\RequirePackage[2020-02-02]{latexrelease}

\documentclass[aps,prl,twocolumn,a4paper,superscriptaddress]{revtex4}

\usepackage[paperwidth=215mm,paperheight=300mm,centering,hmargin=1.6cm,vmargin=2cm]{geometry}
\usepackage{multibbl}
\usepackage{graphicx,amsmath,SIunits}
\usepackage{latexsym,stmaryrd}
\usepackage[sort&compress]{natbib}
\usepackage[dvipsnames]{xcolor}
\bibpunct{[}{]}{,}{n}{}{,}
\usepackage[normalem]{ulem}

\makeatletter
\newcommand*{\citenst}[2][]{%
  \begingroup
  \let\NAT@mbox=\mbox
  \let\@cite\NAT@citenum
  \let\NAT@space\NAT@spacechar
  \let\NAT@super@kern\relax
  \renewcommand\NAT@open{[}%
  \renewcommand\NAT@close{]}%
  \citep{#2}%
  \endgroup
}

\renewcommand{\figurename}{\textbf{Fig.}}

\makeatletter
\renewcommand*{\fnum@figure}{{\normalfont\bfseries \figurename~\thefigure}}
\makeatother

\begin{document}

\title{Cavity optomechanics with Anderson-localized optical modes}

\author{G. Arregui}
\affiliation{Catalan Institute of Nanoscience and Nanotechnology (ICN2), CSIC and The Barcelona Institute of Science and Technology, Campus UAB, Bellaterra, 08193 Barcelona, Spain}
\affiliation{Department of Photonics Engineering, DTU Fotonik, Technical University of Denmark, Building 343, DK-2800 Kgs. Lyngby, Denmark}
\author{R. C. Ng}
\affiliation{Catalan Institute of Nanoscience and Nanotechnology (ICN2), CSIC and The Barcelona Institute of Science and Technology, Campus UAB, Bellaterra, 08193 Barcelona, Spain}
\author{M. Albrechtsen}
\affiliation{Department of Photonics Engineering, DTU Fotonik, Technical University of Denmark, Building 343, DK-2800 Kgs. Lyngby, Denmark}
\author{S. Stobbe}
\affiliation{Department of Photonics Engineering, DTU Fotonik, Technical University of Denmark, Building 343, DK-2800 Kgs. Lyngby, Denmark}
\author{C. M. Sotomayor-Torres}
\affiliation{Catalan Institute of Nanoscience and Nanotechnology (ICN2), CSIC and The Barcelona Institute of Science and Technology, Campus UAB, Bellaterra, 08193 Barcelona, Spain}
\affiliation{ICREA - Instituci\'o Catalana de Recerca i Estudis Avan\c{c}ats, 08010 Barcelona, Spain}
\author{P. D. Garc\'{i}a}
\email{david.garcia@icn2.cat}
\affiliation{Catalan Institute of Nanoscience and Nanotechnology (ICN2), CSIC and The Barcelona Institute of Science and Technology, Campus UAB, Bellaterra, 08193 Barcelona, Spain}
\homepage{http://www.icn.cat/~p2n/}

\date{\today}

\small

\begin{abstract}
Confining photons in cavities enhances the interactions between light and matter.\ In cavity optomechanics, this enables a wealth of phenomena ranging from optomechanically induced transparency to macroscopic objects cooled to their motional ground state.\ Previous work in cavity optomechanics employed devices where ubiquitous structural disorder played no role beyond perturbing resonance frequencies and quality factors.\ More generally, the interplay between disorder, which must be described by statistical physics, and optomechanical effects has thus far been unexplored.\ Here, we demonstrate how sidewall roughness in air-slot photonic-crystal waveguides can induce sufficiently strong backscattering of slot-guided light to create Anderson-localized modes with quality factors as high as half a million and mode volumes that are below the diffraction limit.\ We observe how the interaction between these disorder-induced optical modes and in-plane mechanical modes of the slotted membrane is governed by a distribution of coupling rates, which can exceed $g_{\text{o}}/2\pi\sim  200$ kHz, leading to mechanical amplification up to self sustained oscillations via optomechanical backaction.\ Our work constitutes the first steps towards understanding optomechanics in the multiple-scattering regime and opens new perspectives for exploring complex systems with multitude mutually-coupled degrees of freedom.
\end{abstract}


\maketitle

The presence of thermally excited vibrations in solid-state materials is unavoidable.\ These vibrations modify the interaction between light and matter~\cite{Raman, Brillouin}, which often leads to undesirable effects such as the dephasing of single photons emitted by self-assembled quantum dots~\cite{QDsdephasing} or the optical beam jitter in interferometric gravitational-wave detectors~\cite{gravitational}.\ When properly controlled, this same interaction facilitates groundbreaking applications such as ultrafast laser spectroscopy~\cite{Qswitching} and passive radiative cooling~\cite{cooling}, and can also offer fundamental insight into open challenges such as quantum gravity~\cite{gravity}, among other quantum technological applications~\cite{Martin-Cano}.\ A conventional method to enable such control is through enhancement of the optomechanical interaction by simultaneously confining the electromagnetic and displacement fields within the same carefully nanostructured volume~\cite{CavityOM}.\ In theory, material and geometry provide very fine control over the parameters governing their dynamics.\ In practice, nanoscale optomechanical systems are challenging to realize exactly as designed due to unavoidable fabrication disorder~\cite{Savona_optimization}.\ Geometric imperfections of even only a few nanometers scatter light and are a source of loss, which decreases the temporal confinement of the wave quantified by the quality factor $Q$ of the cavity, a typical figure of merit for describing the interaction with any form of matter.\ Generally, the strategy to reduce such losses is to improve nanofabrication. Despite extensive efforts to develop fabrication methods to meet demanding tolerances at the nanometer scale, thermodynamics sets an unavoidable limit in the interaction between light and matter by imposing a baseline minimum degree of imperfection in any nanoscale system \cite{imperfections}.\

\begin{figure*}
\centering
 \includegraphics[width=\textwidth]{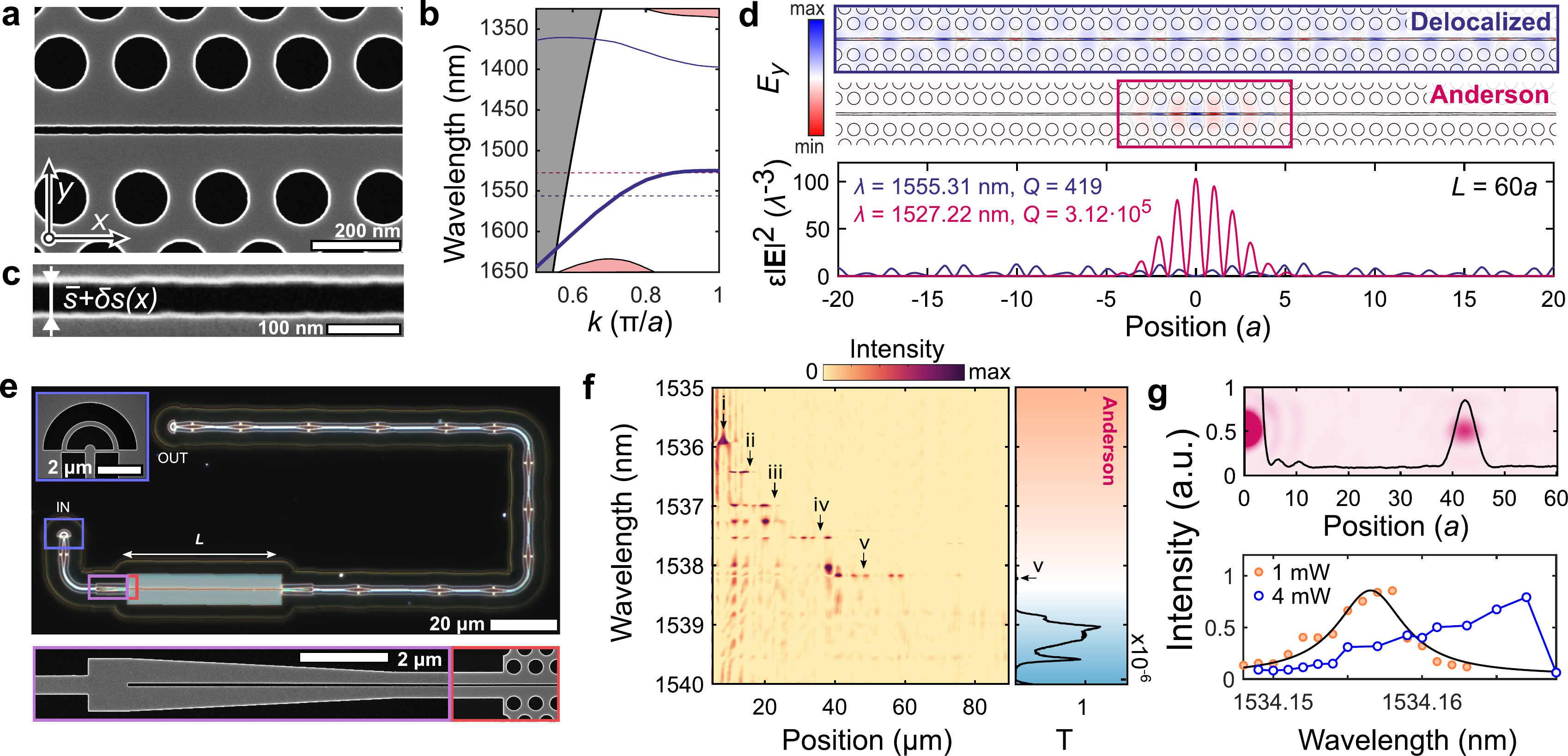}
    \caption{\textbf{Anderson localization of light in air-slot photonic-crystal waveguides.} \textbf{a} Scanning electron microscopy (SEM) micrograph of a slot photonic-crystal waveguide (sPhCW). \textbf{b} Transverse-electric-like band structure of the waveguide shown in \textbf{a}, with parameters: periodicity $a$ = 450 nm, circle radius $r$ = 155 nm, slot width $s$ = 52 nm, and membrane thickness $t$ = 220 nm. \textbf{c} High-magnification SEM focusing on an etched slot of average width $\overline{s}$ = 52 nm and fluctuations $\delta s(x)$ due to line-edge roughness in the sidewalls.\ \textbf{d} Simulated electric-field profiles $E_y(x,y,z=0)$ of a typical delocalized mode (blue box) and an Anderson-localized mode (pink box) in the presence of a 3 nm r.m.s. slot line-edge roughness. The normalized energy density along a cut at the center of the slot is given for both modes, along with the corresponding resonant wavelengths $\lambda$ and quality factors $Q$. \textbf{e} Dark-field microscope image showing a photonic circuit comprised of input and output free-space circular grating couplers (SEM in blue box), strip waveguides, strip-to-slot tapered mode converters (SEM in purple box), slot waveguides, and slot photonic-crystal waveguides (SEM in red box). \textbf{f} Spectro-spatial mapping of the scattered far-fields measured in a $\overline{s}$ = 52 nm, $L$ = 300$a$ sPhCW using a NIR camera. The transmission spectrum measured at the output grating coupler with a single mode fiber is given for reference. \textbf{g} (top) Measured scattered far field of a tightly-localized Anderson mode with estimated mode volume $V_{\text{eff}}\sim \lambda^3/50$ below the diffraction limit and (bottom) extracted intrinsic quality factor $Q_{\text{i}} =2.77\cdot 10^5$ from the intensity at the spatial maxima. The intensity at low input power, $P_{\text{in}}$ = 1 mW, (orange dots) is fitted with a Lorentzian (solid black line) and thermo-optic redshift is observed at higher input power, $P_{\text{in}}$ = 4 mW (blue line).}
    \label{fig:1new}
\end{figure*}

Here, we take an alternative route to cavity optomechanics which deliberately exploits complexity arising from unavoidable surface roughness to mediate the coupling between infrared photons and hypersonic phonons.\ We use a silicon photonic-crystal waveguide to guide light with low group velocity~\cite{Baba} that incorporates an air slot along its length to locally enhance the electromagnetic field within it.\ The subwavelength confinement near the dielectric air-silicon interfaces causes the waveguide mode frequency to be highly dependent on the slot width. In a suspended structure, this enables efficient coupling of the slot-confined light to in-plane mechanical motion, although the local field enhancement also promotes scattering due to surface roughness of the slot sidewalls.\ Beyond these undesired propagation losses~\cite{13}, such nanoscale disorder has more profound consequences such as Anderson localization, an emergent process in which coherent multiple scattering induces spatial confinement of the wave by interference~\cite{Anderson}.\ This localization confines the electromagnetic field to a set of cavity modes at random frequencies and positions along the slot.\ Similar disorder-induced cavities have been observed in conventional photonic-crystal waveguides \cite{topolancik1,topolancik2} and have been used in cavity-quantum-electrodynamics~\cite{Luca,Tyrrestrup}, random lasing~\cite{Jin}, and biosensing~\cite{Vollmer_sensing} experiments.\ The system explored here excels both in terms of the quality factors, $Q$, reaching $5\cdot10^5$, and effective mode volumes, $V_{\text{eff}}$, which are well below the diffraction limit. Such values allow for efficient coupling to the in-plane mechanical motion of the suspended structure in a wide frequency range up to 600 MHz.\ This cavity-optomechanical interaction is enhanced in the slow-light regime of the waveguide~\cite{particle} where light localization occurs, leading to optomechanical coupling strengths as high as $g_{\text{o}}/2\pi\sim  200$ kHz, enabling the observation of dynamical back-action amplification and self-sustained coherent mechanical oscillations, i.e., phonon lasing, at sub-mW optical powers.\

The slot photonic-crystal waveguides (sPhCW) we explore here are suspended line-defect silicon photonic-crystal waveguides with an air slot of width $s$ along the axis of the waveguide, as shown in the scanning electron micrograph of Fig.~\ref{fig:1new}\textbf{a}. Fig.~\ref{fig:1new}\textbf{b} depicts the transverse-electric-like (TE-like) band structure of a waveguide with periodicity $a$ = 450 nm, circle radius $r$ = 155 nm, slot width $s$ = 52 nm, and membrane thickness $t$ = 220 nm. The lowest energy guided mode along the defect (dark blue line in Fig.~\ref{fig:1new}\textbf{b}) exhibits subwavelength light confinement within the air slot while slowing it down with a very large group index, $n_{\text{g}} = c/\left|v_{\text{g}}\right|$, at the Brillouin Zone edge ($k =\pi/a$), see Supplementary Section S1.1 for details. In addition, the waveguide is monomode over its frequency range. The fabricated waveguides are affected by line-edge roughness along the etched silicon sidewalls, which becomes especially prominent for the narrowest-etched features \cite{RIE_lag}. This roughness produces stochastic fluctuations $\delta s(x)$ of the slot width along the waveguide axis, as shown in Fig.~\ref{fig:1new}\textbf{c} for a slot of average width $\overline{s}$ = 52 nm. Such nanoscale roughness leads to uncontrolled extrinsic scattering, which for a monomode waveguide comes only in the form of out-of-plane radiation and backscattering~\cite{lalanne_rho}. Both processes are enhanced when the intensity of the optical mode is high at  air-dielectric interfaces~\cite{ieee}, as is the case at the air-slot boundaries. While the effect of scattering into the radiation continuum can be considered as a sum of incoherent scattering events~\cite{hughes_losses}, coherent multiple backscattering leads to interference effects that spectro-spatially localize the field in a quasi-one-dimensional manifestation of Anderson localization~\cite{Light-matter}. In practice, this localization occurs when the waveguide length, $L$, is larger than the characteristic scattering mean free path, otherwise known as the localization length, $\xi$~\cite{Anderson}, which can be very short for slow light (see Supplementary Section S1.2 for numerical calculations). Disorder-induced optical modes resulting from this localization occur over a narrow wavelength range around the cut-off wavelength, whose extent is determined by $\xi/L$, which in turn depends on the average width of the slot, $\overline{s}$, and the stochastic properties of the line-edge roughness, $\delta s$. The top two panels of Fig.~\ref{fig:1new}\textbf{d} depict the $y$-component of the electromagnetic field for a characteristic delocalized (blue box) and a characteristic Anderson-localized (pink box) mode, and the associated wavelengths of these modes are indicated by horizontal dashed lines in their respective colors in Fig.~\ref{fig:1new}\textbf{b}. These mode profiles are obtained from three-dimensional finite-element simulations of open waveguides of length $L$ = 60$a$ that account for slot line-edge roughness, which we model as normally-distributed fluctuations with r.m.s amplitude $\sigma$ = 3 nm and an exponential correlation function (see Supplementary Section S1.2). Wavelength-scale spatial localization with a single field maxima~\cite{lowerbound} of the Anderson-localized mode occurs in the slow-light region of a slot photonic-crystal waveguide. We compute the effective volume of this mode as $V_{\text{eff}}(\mathbf{r}_{\text{o}}) = 1/\text{Re}\{\epsilon(\textbf{r}_{\text{o}})\textbf{E}(\textbf{r}_{\text{o}})^2\}$, where the location $\textbf{r}_{\text{o}}$ is the position of highest intensity of the mode along the axis of the slot. This corresponds to the inverse of the maximum value of the normalized energy density shown in the bottom panel of Fig.~\ref{fig:1new}\textbf{d}, which suggests effective mode volumes as small as $V_{\text{eff}}\sim\lambda^3/100$, well below the diffraction limit. This tight confinement results from a combination of 1) light being squeezed into the air slot in the propagating Bloch with unit-cell effective mode volumes, $V_{\text{eff,cell}}$ down to $\lambda^3/250$ for $s$ = 52 nm (see Supplementary Section S3.6), and 2) being localized along the waveguide to a length scale given by $\xi$.

\begin{figure}
\centering
 \includegraphics[width=\columnwidth]{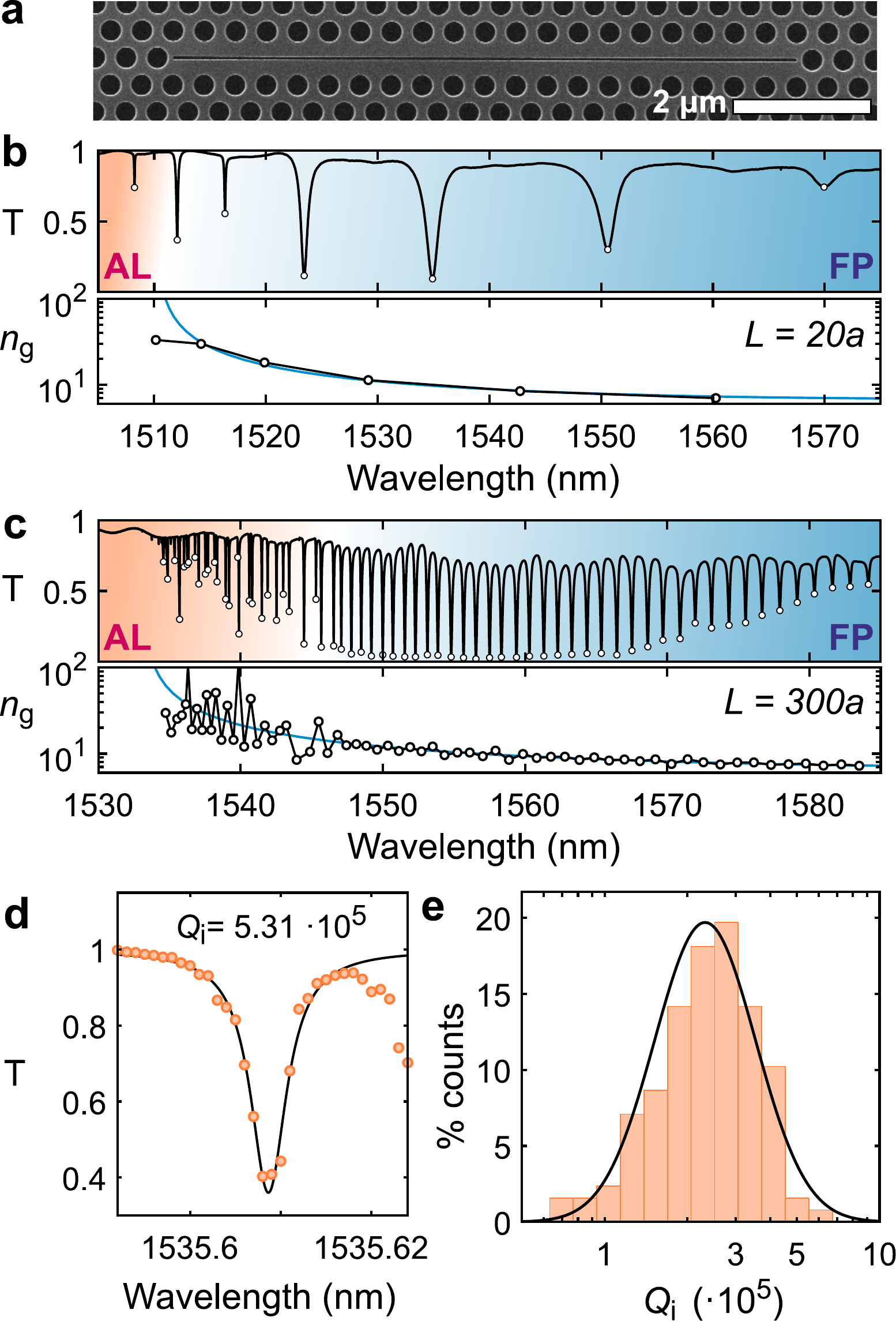}
    \caption{\textbf{Evanescent coupling to Anderson-localized optical modes in slot photonic-crystal waveguides.} \textbf{a} SEM of a slot photonic-crystal waveguide of length $L$ = 20$a$ and slot width $\overline{s}$ = 52 nm, terminated with photonic-crystal mirrors. A tapered fiber loop is placed approximately at the center of the waveguide and is used to evanescently couple into the cavity modes.\textbf{b} Optical transmission spectrum of the structure shown in \textbf{a} and comparison of the reconstructed $n_{\text{g,exp}}(\lambda)$ (solid black line) and simulated $n_{\text{g,sim}}(\lambda)$ group indices (solid blue line). \textbf{c} Same as \textbf{b} for a $L$=300$a$ waveguide. Both \textbf{b} and \textbf{c} exhibit a crossover between ballistic transport forming Fabry-P\'erot resonances (FP, shaded blue) to Anderson-localized modes (AL, shaded orange).\ While the free spectral range varies monotonously in the FP region, it is randomized in the AL region.\ \textbf{d} Optical spectrum of the Anderson mode with the highest intrinsic quality factor, $Q_{\text{i}}$. \textbf{e} Histogram of $Q_{\text{i}}$ for the Anderson-localized modes measured within 11 nominally-identical waveguides of slot width $\overline{s}$ = 52 nm and length $L$ = 300$a$. The solid black line is a fit to a log-normal distribution.}
    \label{fig:2}
\end{figure}

We perform far-field optical measurements of these slot photonic-crystal waveguides by embedding them into photonic circuits, such as the one shown in the dark field optical microscope image of Fig.~\ref{fig:1new}\textbf{e}. For this purpose, we couple TE-like light into a strip waveguide using a broadband grating coupler~\cite{gratingcoupler} (Fig.~\ref{fig:1new}\textbf{e}, blue box), which is then transferred into a slot waveguide with a strip-to-slot mode converter~\cite{striptoslot} (Fig.~\ref{fig:1new}\textbf{e}, purple box) and subsequently interfaced with the slot photonic-crystal waveguide of length $L$ (Fig.~\ref{fig:1new}\textbf{e}, red box). After propagation through the waveguide, the previous optical elements are mirrored. We acquire the scattered far field directly from the circuit using a near-infrared camera. The colormap in Fig.~\ref{fig:1new}\textbf{f} shows the signal obtained along the waveguide axis as a function of wavelength around the cut-off observed in the circuit transmission ($\lambda \sim$ 1538 nm), shown to the right of the colormap. The spectro-spatial map reveals the presence of intense scattered fields around the observed cut-off wavelength. The Lorentzian-shaped peaks in wavelength are evidence of the formation of cavities with a spatial extent that decreases for shorter wavelengths. Several modes are highlighted with arrows. The most extended one, labelled (v), results in a small resonant peak in transmission while more localized modes, (i)-(iv), do not appear in transmission. In this set of waveguides, the most localized mode exhibits a single Gaussian-like hot-spot of width $\sim 4a$ (Fig.~\ref{fig:1new}\textbf{g}) and an intrinsic quality factor $Q_{\text{i}}\sim 2.8\cdot10^5$.\ Such strong spectral and spatial confinement leads to a thermo-optic redshift \cite{Thermo-optic} when the incident power increases.\ Despite the prominent light confinement in the air slot, the high energy density at the sidewalls likely enhances absorption mediated by surface states \cite{SSA}. The effective volume of this mode is estimated to be $V_{\text{eff}}\sim \lambda^3/50$ by assuming that the far field is proportional to the near-field envelope of the cavity mode and by using the value for $V_{\text{eff,cell}}$ given previously (see Supplementary Section S3.1 for details). This estimation agrees well with simulated values of $V_{\text{eff}}$, see Fig.~\ref{fig:1new}\textbf{d}, and leads to a $Q_{\text{i}}/V_{\text{eff}} \sim 1.5\cdot10^7 \lambda^{-3}$, which is among the highest values ever reported for an optical cavity (see recent survey in Ref.~\cite{Q_V}).\\

As evidenced in the spectral map of Fig.~\ref{fig:1new}f, butt-coupling from an integrated access waveguide only allows the excitation of Anderson-localized modes that are relatively close to the input of the waveguides, at a distance approximately given by the localization length $\xi$. To circumvent this limitation and improve our statistical analysis, we employ an alternative near-field technique based on a tapered fiber loop that is placed in contact with the structure along the waveguide axis~\cite{onedimensionalOMC}. In this case, the waveguides are terminated on both ends by a photonic crystal which creates very long cavities (see Fig.~\ref{fig:2}\textbf{a} and Supplementary Section S1.2) that allow for the mapping of their dispersion relation~\cite{lee_characterizing_2008}.\ Evanescent coupling to resonant modes of this cavity-waveguide appear as sharp spectral dips in the transmitted optical signal, as shown in the top panels of Fig.~\ref{fig:2}\textbf{b} and \textbf{c}, which correspond to devices of length $L$ = 20$a$ and $L$ = 300$a$, respectively.\ The strong dispersion of $\xi(\lambda)$ in these systems~\cite{dispersion} effectively leads to two transport regimes.\ In the first regime, where $\xi(\lambda)\gg L$ (blue shaded region in Figs.~\ref{fig:2}\textbf{b} and \textbf{c}), light is largely unaffected by disorder and travels quasi-ballistically all along the waveguide building up Fabry-Pérot modes.\ For the second regime, where $\xi(\lambda)\ll L$ (orange shaded region in Figs.~\ref{fig:2}\textbf{b} and \textbf{c}), backscattering is sufficiently strong to lead to the formation of Anderson-localized modes such as those shown in Fig.~\ref{fig:1new}. These two regimes are directly visible from the optical spectra measured by the fiber. While the Fabry-Pérot modes have resonant wavelengths determined by the waveguide group index $n_{\text{g,exp}}(\lambda)$ and the cavity length $L$, the Anderson-localized modes have randomly distributed resonant wavelengths with strong variations in the coupled fraction of the optical signal, i.e., the depth of the transmission dips. Their localized nature can be confirmed by placing the loop at different positions along the waveguide. We use the free spectral range of all of the observed dips in the Fabry-P\'erot regime to reconstruct the group index of the waveguide mode, $n_{\text{g,exp}}(\lambda)$ (Supplementary Section S3.3 for details). Below a certain value of $n_{\text{g}}$, we observe a very good agreement between the simulated $n_{\text{g,sim}}(\lambda)$ (solid blue line in the bottom panels of Figs~\ref{fig:2}\textbf{b} and \textbf{c}) and $n_{\text{g,exp}}(\lambda)$ (solid black line in the bottom panels of Figs~\ref{fig:2}\textbf{b} and \textbf{c}). Above a certain value of $n_{\text{g}}$, Anderson-localization leads to strong fluctuations of $n_{\text{g,exp}}(\lambda)$ around $n_{\text{g,sim}}(\lambda)$.\ The exact crossover between the two transport regimes depends on the length of the waveguide \cite{chirag}: it occurs at lower $n_{\text{g}}$ for longer waveguides, as shown in Figs.~\ref{fig:2}\textbf{b} and \textbf{c}. The strong backscattering observed at $n_{\text{g}}\sim30$ for a waveguide with $L$ = 20$a$ indicates a very short localization length $\xi$ with a value of $\xi$ that rapidly decreases with increasing group index, as confirmed by the numerical simulations in Supplementary Section S1.2.\\

By using the fiber loop to couple into the waveguide, we observe a much greater number of Anderson-localized modes than in the in-line coupling approach described in Fig.~\ref{fig:1new}, especially for the longest cavity-waveguides ($L$ = 300$a$). Most of these modes appear in a 10 nm range around the cut-off wavelength and have very high quality factors, far beyond previous observations of Anderson-localized modes. The highest intrinsic quality factor, $Q_{\text{i}}$, observed in a waveguide with $\overline{s}$ = 52 nm is above $5\cdot 10^5$, as shown via the optical transmission in Fig.~\ref{fig:2}\textbf{d}. We perform a statistical analysis of the achieved $Q_{\text{i}}$ by placing the fiber loop at the central position of 11 nominally-identical structures, and summarize the results for $\overline{s}$ = 52 nm in the histogram in Fig.~\ref{fig:2}\textbf{e}. Most of the observed modes have $Q_{\text{i}} > 10^{5}$, in agreement with the few modes observed via the far-field measurements of Fig.~\ref{fig:1new}\textbf{f}. The histogram is well fitted with a log-normal distribution, showing the expected behavior deep within the localization regime~\cite{Smolka,Vasco}. We measure similar distributions for all the waveguides with slot widths $\overline{s}>$ 30 nm, with even higher values observed for wider slots (see Supplementary Section S4.5). While the largest $Q_{\text{i}}$ are measured for waveguides of length $L$ = 300$a$, very similar values are also obtained for relatively short structures, indicating that the measured $Q$s are not limited by finite-size effects and only depend on spatially-distributed out-of-plane radiation losses. The ultra-high values of $Q$ observed and the estimated values of the mode volume $V_{\text{eff}}$ are comparable to and even greater than those reported in slot-mode engineered cavities~\cite{ASNaeini,cheeweiwong2,OEMsoren}, where confinement is achieved through carefully tailored potentials. This highlights the potential of such disorder-induced cavities not only for cavity quantum electrodynamics experiments with trapped atoms~\cite{QED_slot}, where $Q_{\text{i}}/V_{\text{eff}}$ would naturally emerge, but also for nonlinear optics~\cite{nonlinearoptics_slot}, sensing~\cite{sensing_slot}, and cavity optomechanics (as shown here).\\

 \begin{figure}
\centering
 \includegraphics[width=\columnwidth]{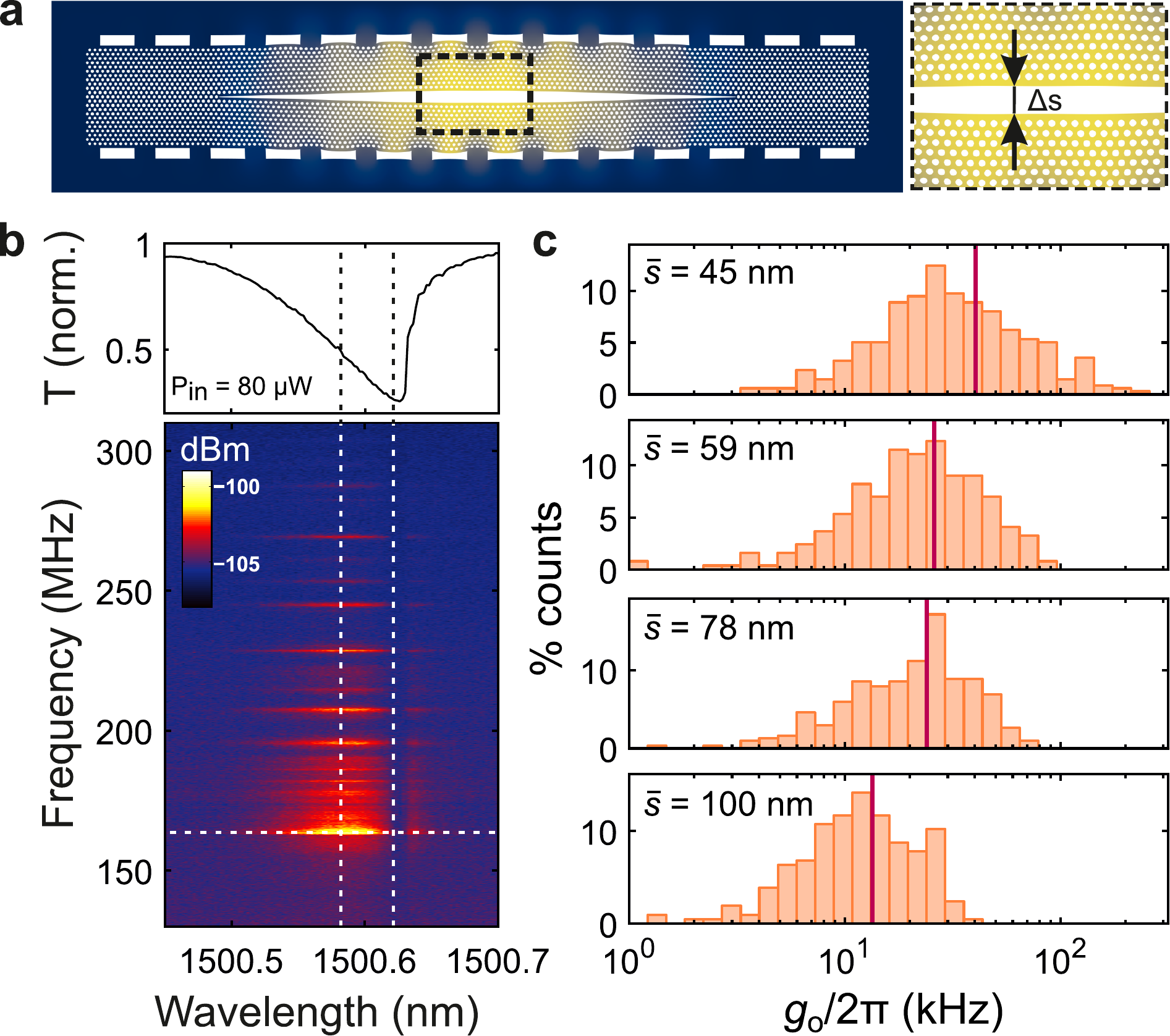}
    \caption{\textbf{Optomechanical coupling in the Anderson localization regime.} \textbf{a} Fundamental in-plane mechanical eigenmode calculated with a 3D finite-element method for a cavity of length $L$ = 50$a$, where $a$ = 470 nm is the lattice separation between circular holes.\ The inset shows the deformation profile in the central region, where the local variation of the slot width $\Delta s$ induced by the mechanical motion is maximal.\ \textbf{b} Direct current value (top) and radiofrequency spectrum (bottom) of the transmitted laser light through a fiber placed in close proximity to an Anderson-localized mode measured while sweeping the input laser source across the optical resonance.\ Here the cavity length is $L$ = 145 $\mu$m and $\overline{s}$ = 78 nm.\ The thermally active mechanical modes coupled to the cavity appear as horizontal lines and have Lorentzian lineshapes.\ The fundamental mode is highlighted with a horizontal dashed line.\ The vertical dashed lines across both panels identify the points of maximum and vanishing transduction.\ \textbf{c} Log-scale histogram of the vacuum optomechanical coupling rate $g_{\text{o}}$ obtained from multiple mechanical and Anderson-localized optical pairs in four different waveguides with different slot widths $\overline{s}$.\ The average value in each histogram is indicated with a solid vertical line.}
    \label{fig:3}
\end{figure}

 \begin{figure*}
\centering
 \includegraphics[width=\textwidth]{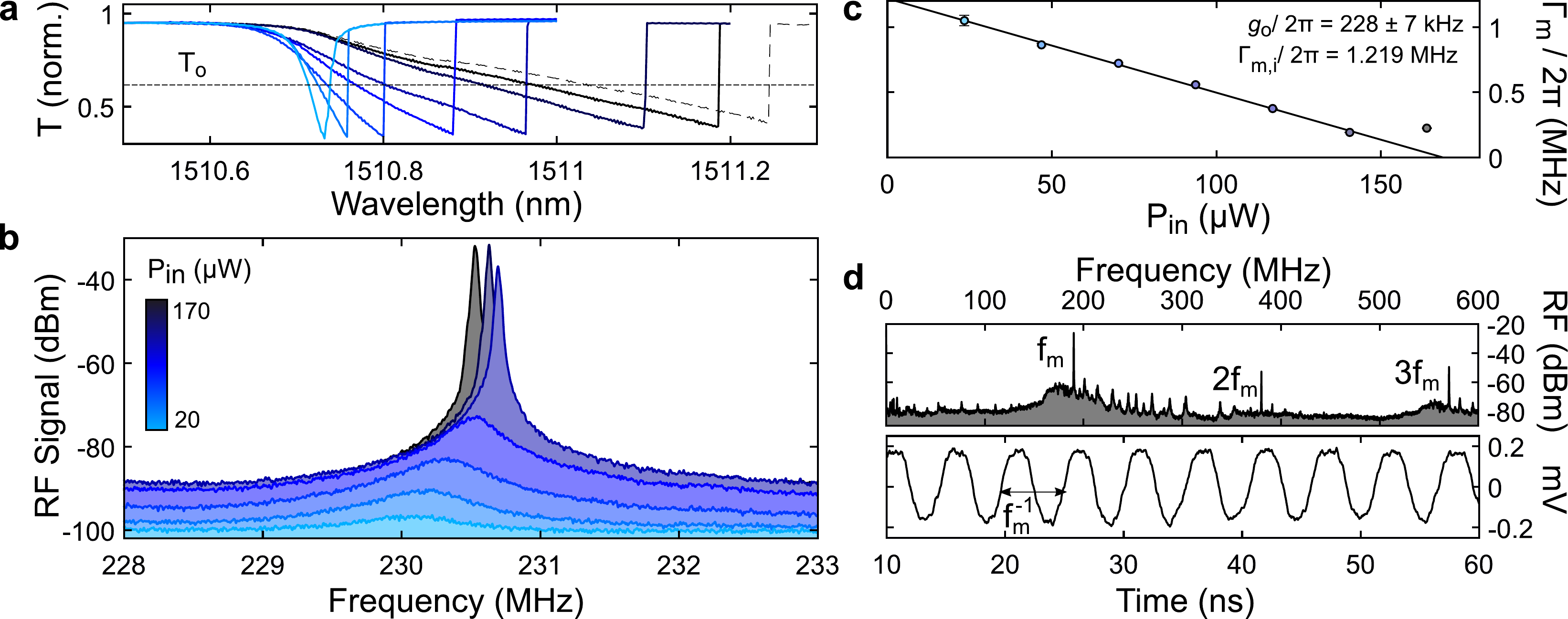}
    \caption{\textbf{Optomechanical amplification and phonon lasing.} \textbf{a} Optical transmission spectrum across an Anderson-localized cavity at various input excitation powers $P_{\text{in}}$.\ The dashed line indicates the coupled fraction at which the various radiofrequency spectra in \textbf{b} are acquired.\ The color scale indicates the input excitation power and is consistent throughout the figure.\ \textbf{c} Effective mechanical linewidth $\Gamma_{\text{m}}^{\text{eff}}$ measured as a function of $P_{\text{in}}$ (blue dots) along with a linear fit (solid black line).\ The slope is used to extract the $g_{\text{o}}$ value as $\Gamma_{\text{m}}^{\text{eff}} \propto {g_{\text{o}}}^2$.\ For $P_{\text{in}} >$ 150 {\micro\watt}, the mechanical linewidth saturates above the lasing threshold for which the system reaches phonon lasing indicated by the grey area.\ \textbf{d} Characteristic radiofrequency spectrum (top panel) and time trace (bottom panel) of the optomechanical system undergoing coherent self-sustained oscillations, i.e., phonon lasing.}
    \label{fig:4}
\end{figure*}

The strong dependence of the slot-guided light on the exact topography of the sidewalls, which promotes Anderson localization of light, also makes the disorder-induced cavities excellent optical probes of the mechanical vibrational state of the suspended structure.\ The frequency of the confined electromagnetic field within the subwavelength air slot is extremely sensitive to slot size variations and therefore allows precise sensing of mechanical displacement of the nanostructure~\cite{zipper,SeidlerIBM,L3femtogram}.\ In Fig.~\ref{fig:3}\textbf{a}, we plot the typical displacement profile of the fundamental in-plane mechanical mode of the slotted membrane, which includes a series of etched trenches on both sides.\ The inset (dotted black box) highlights the local variation in the slot width, $\Delta\overline{s}$, which modulates the resonance frequency of the optical Fabry-Pérot and Anderson cavity modes of the structure.\ To probe this mechanical modulation, we drive the (Anderson) cavity modes at moderate powers $P_{\text{in}}\sim$ 10 to 100 $\mu$W.\ An example of the transmitted optical signal through one of these cavities is plotted in the upper panel of Fig.~\ref{fig:3}\textbf{b}, measured while scanning the laser wavelength across the optical resonance.\  As already observed in the circuit-embedded waveguides in Fig. \ref{fig:1new}\textbf{e}, the departure from a Lorentzian-lineshape results from the thermo-optic effect. The lower panel of Fig.~\ref{fig:3}\textbf{b} (colormap) shows the transduced radiofrequency modulation of this particular Anderson mode, where each of the horizontal Lorentzian peaks corresponds to a thermally active mechanical eigenmode of the structure, all with mechanical linewidths of $\Gamma_{\text{m}}/2\pi\sim$ 1 MHz.\ The most intense peak in the spectrum (indicated by the horizontal white-dashed line) has a frequency of $\Omega_{\text{m}}/2\pi$ = 161 MHz and corresponds, for the measured cavity length, to the fundamental in-plane mechanical mode plotted in Fig.~\ref{fig:3}\textbf{a} (see Supplementary Section S4).\ In this spectral scan, the two vertical white-dashed lines highlight wavelengths or relative cavity-pump laser detunings, $\Delta$, at which the mechanical transduction is either maximized ($\Delta=\kappa/(2\sqrt{3})$) or vanishes ($\Delta=0$).\ The former corresponds to the maximum derivative of the optical mode transmission and the latter to the resonance wavelength itself.\ The amplitude of the mechanically-induced radiofrequency components across the optical resonance indicate a dispersive optomechanical coupling between the probed optical mode and the mechanical modes.\ We can therefore quantify the photon-phonon coupling by the vacuum optomechanical coupling rate, $g_{\text{o}}$, which can be measured by driving the optical cavity with a known phase-modulated signal~\cite{gomcalib} (see Supplementary Section S5).\ We use this calibration to measure the statistical distribution of $g_{\text{o}}$ between all the detected Anderson-localized modes and all the transduced mechanical modes for different values of $\overline{s}$, plotted in Fig.~\ref{fig:3}\textbf{c}.\ The average value of $g_{\text{o}}$ (solid vertical line) increases for narrower air slots.\ This trend is expected for an engineered adiabatic slot nanocavity~\cite{VerhagenNatComm,optica} where $g_{\text{o}}$ can be quite accurately estimated from the field envelope and the unit cell optomechanical coupling, $g_{\text{o,cell}}$ \cite{safavi_design}. Since the envelope, i.e., the spatial extent of the cavity mode(s) along the waveguide, is roughly independent of $\overline{s}$, the dependence of the band structure on $s$ is sufficient to understand the behavior. \ Here we observe the same average trend even in the case where the field extension along the waveguide, which is statistically given by $\xi$, may strongly depend on $\overline{s}$. This can occur either through $\overline{s}$-dependent scattering cross sections, as expected from first-order perturbation theory \cite{hughes_losses}, or due to enhanced line-edge roughness for smaller widths, which we observe from SEM images. However, the fast drop of the localization length $\xi$ with $n_{\text{g}}$ (see Supplementary Section S1.2) limits the practical differences in $\xi$ for the probed high-$n_{\text{g}}$ region. We therefore attribute the observed trend in Fig.~\ref{fig:3}\textbf{c} to that of the band structure itself. The larger drop observed between $\overline{s}=78 nm$ and $\overline{s}=100 nm$ likely originates from a less efficient localization. This would result from coupling/scattering into a nearly-degenerate transverse-magnetic-like band induced by vertical sidewalls (see Supplementary Sections S2 and S4.5). \\

The high values of $Q_{\text{i}}$ and $g_{\text{o}}$ measured in our system allow for the observation of dynamical back-action in the Anderson localization regime. The disorder-induced optical modes can be used to control the vibrational amplitude and frequency of the mechanical modes through radiation pressure.\ Exploiting such force enables one to amplify (heat) or damp (cool) a mechanical eigenmode  of the system by driving an optical mode with a blue ($\Delta>0$) or red-detuned ($\Delta<0$) laser, respectively.\ The latter is not achieved in our system due to enhanced thermo-optic redshifting with increasing power~\cite{Nonlinear-Anderson}.\ This is shown in Fig.~\ref{fig:4}\textbf{a}, which plots the optical transmission through an Anderson-localized mode when increasing the excitation laser power, $P_{\text{in}}$.\ In addition, the thermo-optic shift complicates the precise experimental determination and locking of the detuning $\Delta$. Here, we assume that a value of $\Delta$ corresponds to a fixed coupled fraction, $T_{\text{o}}$, indicated by the horizontal dashed line in Fig.~\ref{fig:4}\textbf{a}, an approach which neglects any non-linear optical losses. By taking radiofrequency spectra at a fixed $T_{\text{o}}$ and varying power, we observe mechanical amplification through dynamical back-action, as demonstrated by the spectra of a single mechanical resonance with natural frequency $\Omega_{\text{m}}/2\pi$ = 230 MHz (Fig.~\ref{fig:4}\textbf{b}). We observe a frequency blueshift of the mechanical mode, a decrease in the mechanical linewidth, and an increase in the mechanical amplitude.\ Assuming the optomechanical damping,  $\delta\Gamma_{\text{om}}$, is the only source of linewidth narrowing, i.e., $\Gamma_{\text{m}}^{\text{eff}} = \Gamma_{\text{m}}+\delta\Gamma_{\text{om}}$, we expect a linear dependence of the mechanical linewidth with $P_{\text{in}}$, where the slope is proportional to the optomechanical coupling $g_{\text{o}}^2$ \cite{CavityOM}.\ This dependency is confirmed in Fig.~\ref{fig:4}\textbf{c} and provides an alternative method to extract the experimental optomechanical coupling, which is found to be $g_{\text{o}}/2\pi=228\pm7$ kHz for this particular pair of optical and mechanical modes.\ This value is slightly larger than the one obtained using the phase-modulation technique, which indicates that other physical processes may also contribute to the effective optomechanical back-action~\cite{zipper,optica}.\\

When $\delta\Gamma_{\text{om}}\approx-\Gamma_{\text{m}}$, the optomechanical system reaches a regime of regenerative self-sustained oscillations or mechanical lasing, where $\Gamma_{\text{m}}^{\text{eff}}$ saturates \cite{vahala}.\ We observe this saturation above $P_{\text{in}} \sim 150$ {\micro\watt}, highlighted by the grey area in Fig.~\ref{fig:3}c.\ Passive mechanical losses are compensated and the overall mechanical damping rate becomes negative, $\Gamma_{\text{m}}^{\text{eff}}<0$. The mechanical vibration undergoes a transition from thermal Brownian motion into self-sustained monochromatic coherent mechanical oscillations.\ Fig.~\ref{fig:4}\textbf{d} plots the coherently amplified mechanical oscillation in the frequency domain (top panel) and in the time domain (lower panel) induced by the dynamical back-action of an Anderson-localized mode at an excitation power of $P_{\text{in}}$ = 200 {\micro\watt}, above the lasing threshold.\ In the frequency domain, the wide radiofrequency spectrum exhibits multiple high-amplitude harmonics of the lasing mode. The temporal trace plotted in the lower panel of Fig.~\ref{fig:4}\textbf{d} indicates its coherent and quasi-harmonic nature.\ The mechanical frequency of the lasing mode is $\Omega_{\text{m}}/2\pi$ = 190 MHz, which differs from the mode amplified in Fig.~\ref{fig:4}\textbf{b} at lower powers.\ While the optical mode is selected by the laser drive, the mechanical modes coupled to the driven cavity are in the sideband unresolved regime ($\Omega_{\text{m}}/\kappa = 10^{-2}-10^{-1}$) and compete for the same optomechanical gain.\ The multiple mechanical modes observed here (see Fig. \ref{fig:3}\textbf{b}) have comparable intrinsic mechanical linewidths $\Gamma_{\text{m}}$ and optomechanical couplings $g_{\text{o}}$ to those of the driven optical cavity, which sets their mechanical lasing threshold at similar values. Phonon lasing initially takes place at the mode with the lowest threshold (Fig.~\ref{fig:4}\textbf{b}), whilst the other modes are cooled~\cite{anomalouscooling}.\ On the contrary, when the pump power is considerably larger than the threshold for two or more modes, that with the largest radiation-pressure coupling $g_{\text{o}}$ is finally amplified (Fig.~\ref{fig:4}\textbf{d}).\\

In recent years, well-known functionalities such as  cavity quantum electrodynamics~\cite{Luca}, lasing~\cite{Jin}, imaging~\cite{Bertolotti}, and computing~\cite{Sylvain} have been explored in novel systems where complexity in the form of structural imperfection plays a primary role.\ Here, we exploited disorder to mediate the optomechanical coupling between infrared photons and radiofrequency phonons.\ We used the roughness in a slow-light air-slot photonic-crystal waveguide to confine light, creating a large set of optical cavities with high quality factors and small mode volumes that couple to and amplify the mechanical eigenmodes of the system up to self-sustained oscillations.\ From a practical perspective, the optical modes explored here and their coupling to motion will help elucidate the role of surface roughness in nano-opto-electro-mechanical systems (NOEMS) that employ deep subwavelength air slots to control the flow of light. From a more fundamental perspective, our work constitutes a step forward in unraveling the intricate connection between optical pressure and light transport in the multiple-scattering regime~\cite{OMdisorder}, the physics of which will dominate as photonic nanocavity research approaches length scales on the order of the roughness~\cite{marcus}. The system explored here, with a multitude of mutually-coupled optical and mechanical modes, falls deep within the class of multimode optomechanical systems.\ Consequently, it is a versatile platform to study mechanical mode competition leading to mode hopping~\cite{modehopping} and anomalous cooling~\cite{anomalouscooling}, many-mode phonon lasing~\cite{Mercade}, and cascaded mechanical state transfer~\cite{weaverNATCOM}.\ Owing to the unprecedented light-matter interaction figure of merit estimated in our system, of at least $Q_{\text{i}}/V_{\text{eff}} \sim 1.5\cdot10^7 \lambda^{-3}$ but probably much higher, and the air-confined nature of the involved optical modes, the platform also opens the possibility to explore highly efficient optical excitations with electron beams~\cite{abajo} as well as photon-phonon interactions mediated by trapped atoms~\cite{kimblePNAS}. Finally, while the mechanical modes explored here are essentially unaffected by structural disorder, simultaneous waveguiding of disorder-sensitive GHz slow sound~\cite{Omar,safaviPRL} and telecom slow light is possible in a slotted optomechanical waveguide, enabling the study of acoustic Anderson localization by leveraging the reported optical cavities as efficient transducers~\cite{OMAL}.\ As such, the work presented here is a nascent experimental effort to explore localization phenomena with coupled excitations~\cite{roqueAL}.

\section{Acknowledgements}

This work was supported by the Spanish Minister of Science, Innovation and Universities via the Severo Ochoa Program (Grant No. SEV-2017-0706) as well as by the CERCA Program/Generalitat de Catalunya and the EU-H2020 FET Proactive project TOCHA (No. 824140).\ G.A. was supported by the project RTI2018-093921-A-C44 (SMOOTH). R.C.N. acknowledges funding from the EU-H2020 research and innovation programme under the Marie Sklodowska-Curie (Grant No. 897148). P.D. gratefully acknowledges the support of a Ramon y Cajal fellowship (RYC-2015-18124) and C.M.S.T acknowledges the support of the national project SIP (Grant No. PGC2018-104333-BI00). M.A. and S.S. gratefully acknowledge funding from the Villum Foundation Young Investigator Program (Grant No. 13170), the Danish National Research Foundation (Grant No. DNRF147 - NanoPhoton), Innovation Fund Denmark (Grant No. 0175-00022 - NEXUS), and Independent Research Fund Denmark (Grant No. 0135-00315 - VAFL).

\section{Data availability statement}
The datasets generated during and/or analysed during the current study are available from the corresponding author on reasonable request.


\begin{widetext}

\newpage

\renewcommand{\figurename}{\textbf{Supplementary Figure}}
\makeatletter
\renewcommand{\thefigure}{S\@arabic\c@figure}
\makeatother
\renewcommand\theequation{S\arabic{equation}}
\renewcommand\thetable{S\arabic{table}}
\renewcommand{\bibname}{References}

\renewcommand{\topfraction}{.98}
\renewcommand{\bottomfraction}{.98}

\renewcommand{\figurename}{\textbf{Supplementary Figure}}

\makeatletter
\renewcommand*{\fnum@figure}{{\normalfont\bfseries \figurename~\thefigure}}
\renewcommand{\thefigure}{S\@arabic\c@figure}
\makeatother

\renewcommand\theequation{S\arabic{equation}}

\renewcommand\thetable{S\arabic{table}}

\renewcommand{\bibname}{References}

\renewcommand{\thesection}{S\arabic{section}}

\section*{SUPPLEMENTARY INFORMATION}

\setcounter{figure}{0}
\setcounter{equation}{0}

\normalsize 

\section{S1. Slot-mode: photonic band structure and localization} \label{sec:slotmodebandstructure}

\subsection{S1.1. Photonic band structure}
\label{subsec:sPhCW_bands}

A slotted photonic-crystal waveguide (sPhCW) is obtained by etching an air slot of width $s$ along the axis of a standard (W1) photonic-crystal waveguide (PhCW). A W1 PhCW is formed by leaving out a row of holes in a triangular lattice of circular holes etched on a membrane or slab of a high-refractive index material (e.g. silicon, in our case). Fig.~\ref{fig:sPhCWs}a shows a schematic of the final geometry with a slot of width $s$ = 78 nm, circular hole radii $r$ = 147 nm, pitch $a$ = 470 nm, and thickness $t$ = 240 nm. The photonic-crystal lattices surrounding the missing row exhibit a band gap for TE-like modes, modes with even vector symmetry with respect to the mid-plane of the silicon membrane, such that the waveguide acts as a deterministic line defect inside an otherwise perfect lattice. The TE-like band structure of the sPhCW is shown in Fig.~\ref{fig:sPhCWs}b, where the bulk crystal states are shaded in red and the continuum of radiation states above the light line is shaded in grey. The sPhCW supports three defect-guided modes within the gap and below the light line, labelled A, B, and C in Fig.~\ref{fig:sPhCWs}b. Two of these modes (A and C) have a strong local field enhancement within the air slot, as shown by the field profiles $E_y (x,y,z=0)$ at the high-symmetry point $k=\pi/a$ (Fig.~\ref{fig:sPhCWs}c). The mode profiles also extend partially into the surrounding crystal cladding, leading to a zero group velocity $v_{\text{g}}=\frac{\partial\omega}{\partial k}$ and diverging group index $n_{\text{g}}=c/v_{\text{g}}$ at the Brillouin zone (BZ) edge and allows slot-guided slow light. We denote bands A and C as the fundamental and first-order slot bands, respectively. On the contrary, mode B looks very similar to the first-order y-odd mode of a W1 PhCW \cite{Baba}, and we denote it as the W1-like mode. In this work, we have focused on band A. The waveguide does not sustain any other guided or quasi-guided modes within its frequency range, i.e. it is monomode, which minimises extrinsic out-of-plane losses due to fabrication disorder~\cite{savona_electromagnetic_2011,safavi-naeini_optomechanics_2010}.\\

\begin{figure}[b!]
 \centering
  \includegraphics[width=0.8\columnwidth]{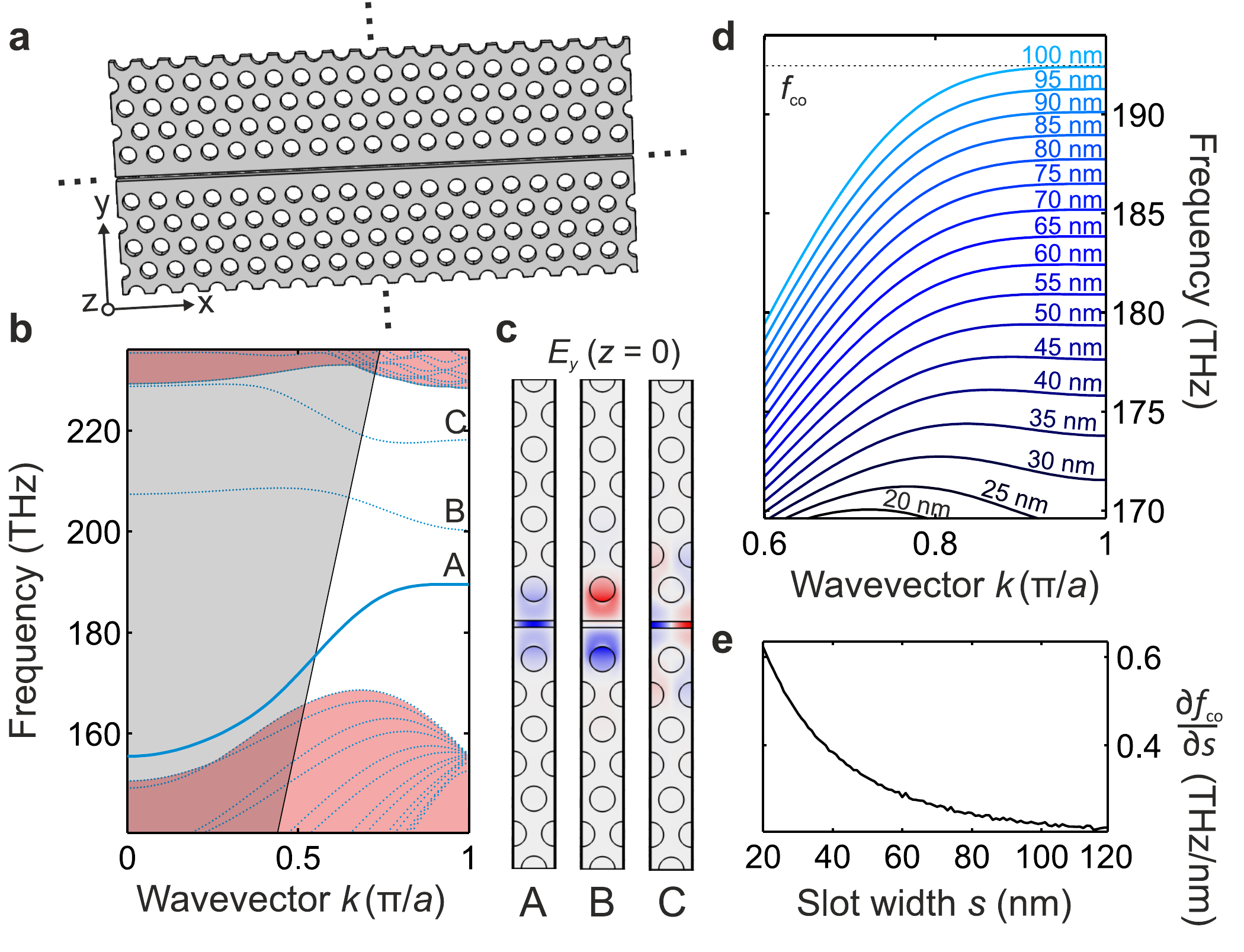}
\caption[Photonic band diagram of a slotted PhCW (sPhCW)]{\textbf{Band structure of a slotted photonic crystal waveguide (sPhCW).} \textbf{a}, Schematic of a sPhCW, which consists of a W1 waveguide with an additional air slot in the direction of the line-defect. \textbf{b}, Band structure of the sPhCW structure shown in (a) with geometrical parameters $s$ = 78 nm, $r$ = 147 nm, $t$ = 240 nm and $a$ = 470 nm. The light cone and the bulk bands are shown with shaded grey and red, respectively. \textbf{c}, $E_y(x,y,z=0)$ for the highlighted bands (A, B and C) at the band edge ($k=\frac{\pi}{a}$), showing either even or odd parity with respect to the waveguide axis. \textbf{d}, Fundamental slot band as a function of the slot width $s$ from $s$ = 20 nm (black) to $s$ = 100 nm (light blue). The cut-off $f_{co}$ frequency is highlighted for the widest slot. \textbf{e}, Rate of change of $f_{co}$ as a function of the slot width $s$, showing an exponential behaviour.}
\label{fig:sPhCWs}
\end{figure}

Due to strong confinement of the light field inside the air slot and its polarization, the frequency of the fundamental slot band (A) is extremely sensitive to the slot width $s$. Fig.~\ref{fig:sPhCWs}d plots the band dispersion as a function of $s$. We calculate a shift of the cut-off frequency, $f_{\text{co}}$, of more than 20 THz when varying the slot size from $s$ = 20 nm to $s$ = 100 nm. The overall dispersion conserves the monomode condition for all slots above $s$ = 40 nm. Fig.~\ref{fig:sPhCWs}e depicts the rate of change of $f_{\text{co}}$, which increases exponentially with decreasing slot size $s$. This high spectral sensitivity, $\Delta f_{\text{co}}/\Delta s$, has a strong impact on the vacuum optomechanical coupling rate, $g_{\text{o}}$, between any cavity mode built from this band and any mechanical mode with an associated change $\Delta s$ of the slot width. Since the optomechanical coupling, $g_{\text{o}}$, quantifies the frequency shift of an optical cavity mode with respect to a displacement of amplitude $x_{\text{zpf}}$, the zero-point fluctuations of the mechanical resonator \cite{CavityOM}, one can also expect an exponential dependence of $g_{\text{o}}$ as a function of $s$~\cite{Leijssen}.\\

\subsection{S1.2. Disorder-induced localization}
\label{subsec:loc_sLN}

Here, we analyze numerically the effect of disorder on the photonic slot-guided mode. In theory, the fundamental mode (A) shown in Fig.~\ref{fig:sPhCWs}b is lossless in the region of interest, i.e., below the light line, and allows $v_{\text{g}}$ tending to zero. In practice, any deviation from the perfect nominal design in the form of structural disorder induces light scattering of the propagating mode in the forward and backward direction into isofrequency modes, i.e., Bloch modes with the same frequency $\omega$~\cite{sheng_introduction_2006}, causing them to interfere with one another. Such interference leads to spatial localization of the light field~\cite{topolancik_experimental_2007} which is a photonic manifestation of Anderson localization~\cite{anderson_absence_1958} in low dimensions. Strong localization occurs when the length of the waveguide $L$ is larger than the characteristic disorder length-scale, the localization length $\xi$, which is a relevant parameter for understanding the physics of a real photonic waveguide. An additional loss channel that need to be considered is the out-of-plane radiation. Although the modes considered here are below the light line in the ideal case and do not suffer from intrinsic out-of plane losses, disorder may couple them to the radiation continuum. For large values of the group index $n_{\text{g}}$ and low disorder levels, in-plane backscattering is typically the dominant loss mechanism in standard W1 waveguides~\cite{savona_electromagnetic_2011}. To verify that this holds true for sPchWs, we calculate both the average transmission decay and their eigenmodes in the presence of fabrication disorder. Due to the slot-confined nature of the field (Fig.~\ref{fig:sPhCWs}c), we only consider line-edge roughness $\delta s(x)$ in the slot sidewalls with a mean width $\overline{s}$. $\delta s(x)$ is modeled independently in the two sidewalls by considering normally distributed fluctuations of zero mean, standard deviation $\sigma$, and exponential autocorrelation function with correlation length $L_c$. A single-slot geometry is built by generating two realisations of $\delta s(x)$ and shifting each of them by $+\overline{s}/2$ and $-\overline{s}/2$, respectively. Although, in principle, both $\sigma$ and $L_c$ can be estimated from SEM micrographs of the samples \cite{houdre}, here we set a value for $\sigma$ (2 or 3 nm) and $L_c$ = 1 $\mu$m. An example of the resulting slot geometry over a length $20a$ is shown in Fig.~\ref{fig:LocsPhCW}a.\\

\begin{figure}[t!]
\centering
\includegraphics[width=0.8\textwidth]{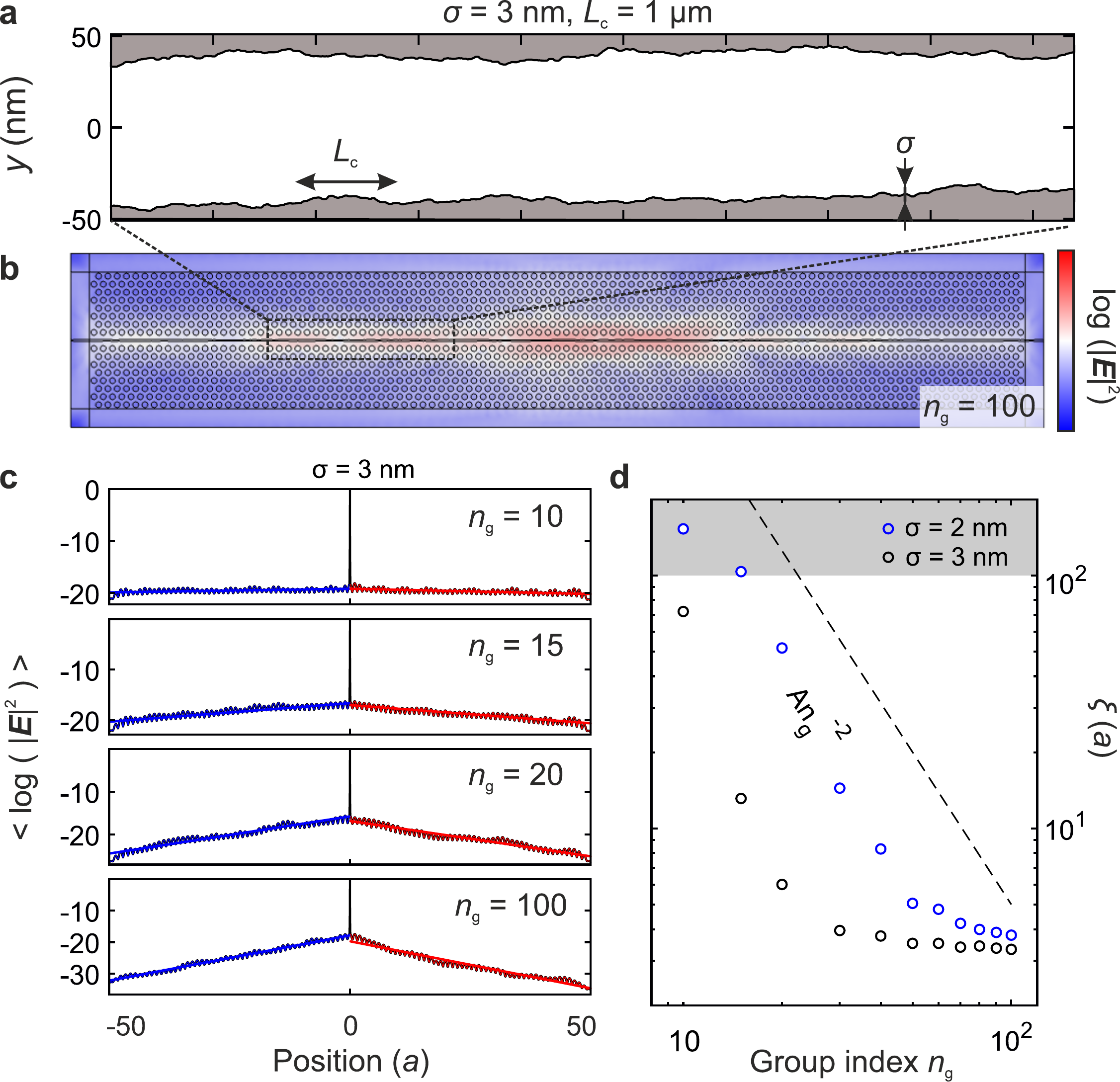}
\caption[Localization length in a sPhCW waveguide]{\textbf{Photonic localization length $\xi$ in a slotted photonic crystal waveguide (sPhCW) in the presence of roughness.} The roughness in the slot sidewalls is modeled as line-edge roughness with r.m.s. amplitude $\sigma$ and correlation length $L_{\text{c}}$. \textbf{a}, Example of a generated slot with $\sigma$ = 3 nm and $L_{\text{c}}$ = 1 $\mu$m. \textbf{b}, Logarithm of the electric field intensity $\lvert E\rvert^2$ emitted by a dipole along the axis of a disordered sPhCW at a frequency corresponding to $n_g$ = 100. \textbf{c}, Ensemble average of the emitted field at 4 different group indices for $\sigma$ = 3 nm. Both sides are linearly fitted (blue and red) and the localization length $\xi$ is extracted as the mean between the two fitted slopes. \textbf{d}, Group-index dependence of the localization length $\xi(n_{\text{g}})$ for $\sigma$ = 2 nm and 3 nm.. The region where the fields decay less than a factor $e^{-1}$ is marked with a grey shading and the dashed line shows the perturbative~\cite{hughes_extrinsic_2005} slope $n_{\text{g}}^{-2}$.}
\label{fig:LocsPhCW}
\end{figure}

\subsubsection{Localization length}
\label{subsubsec:loclength}

The localization length $\xi(\nu)$ is obtained from fully three-dimensional numerical simulations of the light emitted by a light source embedded in the slot, following Ref.~\cite{garcia_two_2017} as
\begin{equation}
\label{eq:xidecay}
-\frac{x}{\xi(\nu)}=<\text{ln}[I(\nu)]>
\end{equation}
where $I=\lvert E \rvert^2$ is the FEM solution of the electromagnetic field intensity emitted by a single $y$-oriented dipole at frequency $\nu$, $x$ is the distance from the dipole position along the waveguide and the brackets indicate the statistical ensemble average over different configurations of disorder. The emitter is placed at the center of a waveguide of length $L=100a$ in the $x$-direction with eight unit cells on each side of the waveguide in the $y$-direction. This domain is surrounded by perfectly matched layers (PMLs) that include a slot continuation to mimic an open system. Note that the assumed exponential damping~\eqref{eq:xidecay} is not found in all types of disordered monomode waveguide systems~\cite{baron_attenuation_2011}, especially as $v_g$ decreases, but it is commonly assumed and recovered numerically~\cite{garcia_two_2017}. Fig.~\ref{fig:LocsPhCW}b plots the electromagnetic field intensity (log-scale) excited by a dipole source at frequency $\nu=187$ THz, or $n_{\text{g}}$ = 100, in a single realisation of disorder ($\sigma$ = 3 nm). The exponential decay is obtained by extracting the field intensity along the axis of the slot and ensemble averaging the emitted fields over 50 disorder realisations, for which the extracted parameter $\xi$ practically converges. The exponential fits to four different group indices are shown by red and blue lines in Fig.~\ref{fig:LocsPhCW}c, indicating well-behaved exponential decays with decreasing characteristic length-scale $\xi$. Taking $\xi$ as the mean between the two fitted lengths, we extract the frequency-dependent $\xi(\nu)$, which is given in Fig.~\ref{fig:LocsPhCW}d  for two different r.m.s disorder levels, with the group index on the x-axis. The shaded region marks the region where the fields decay less than a factor $e^{-1}$ within the simulated length $L$ and the dashed line shows the slope expected from first order perturbation theory $n_g^{-2}$~\cite{hughes_extrinsic_2005}. The decrease of $\xi$ is very pronounced over a short range of $n_{g}$, with $n_g$ values larger than 30 already leading to $\xi<20a$. As seen on the next subsection, this implies tightly localized modes for sPhCWs of much shorter length than those explored in the main text ($\sim 300a$). \\

\subsubsection{Anderson-localized cavity modes}
\label{subsubsec:loc_sLN}

\begin{figure}[t]
\centering
\includegraphics[width=\textwidth]{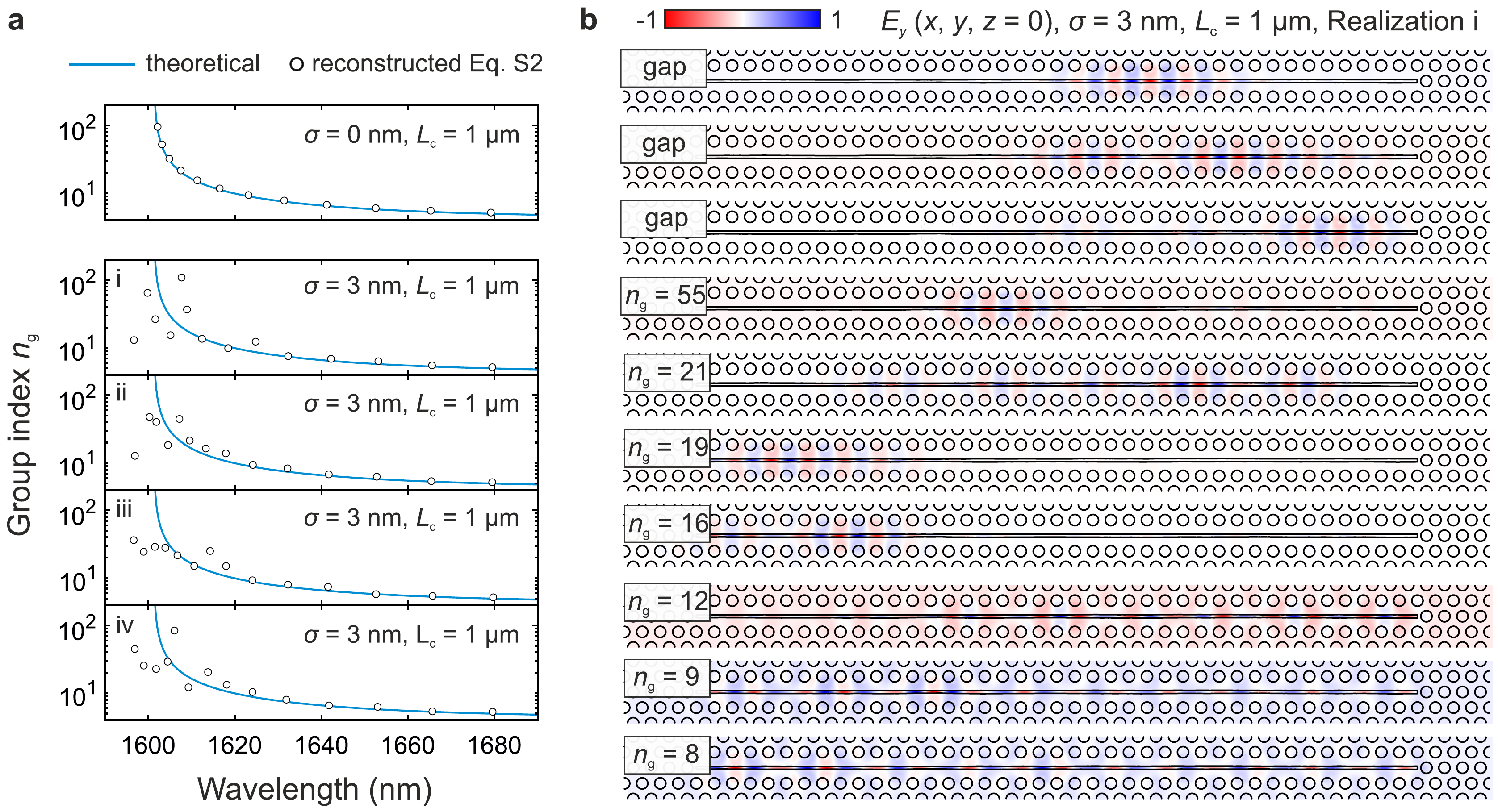}
\caption[Eigenmodes in a disordered sLN cavity]{\textbf{Anderson localization in a slotted photonic-crystal closed waveguide.} The quasi-normal modes (QNMs) of sPhCWs of length $L=40a$ enclosed by mirrors and forming sL40 cavities are simulated by 3D-FEM. Disorder is introduced as line-edge roughness. \textbf{a}, Group index $n_{\text{g}}^{\text{sL40}}$ recovered from the corresponding eigenfrequencies (white dots) for: (top) an unperturbed sL40 cavity and 4 different disorder realisations (bottom panels i-iv). The theoretical group index $n^{\text{sPhCW}}_{\text{g}}$ obtained from the band structure is shown with a solid blue line. \textbf{b}, $E_y(x,y,z=0)$ of the first 10 eigenmodes of the disorder realisation labeled "i" in \textbf{a} with the group index associated with the unperturbed structure shown as indicated in the plot. The label "gap" implies that the mode occurs at a frequency lying within the photonic gap of the waveguide.}
\label{fig:sLNloc}
\end{figure}

When $L$ is larger than the localization length $\xi$, $L\gg\xi(\nu)$, the interference of the ideal Bloch mode leads to a set of localized modes with an average length-scale given by $\xi$~\cite{vasco_statistics_2017}. In standard photonic-crystal waveguides, $\xi(\nu)$ is extremely dispersive due to its relation with the DOS of the system, as plotted in Fig.~\ref{fig:LocsPhCW}d, so a waveguide of fixed length may localize the electromagnetic field at some frequencies while allowing free propagation at others, as shown for an open waveguide in Fig.~1d of the main text. To determine the crossover between the ballistic and the localized regime, we enclose the waveguide with two high-reflectivity mirrors at both extremes, thereby creating a long cavity with dispersive group index which induces cavity Fabry-P\'{e}rot (FP) resonances with varying free spectral range (FSR). By measuring the FSR, we can determine whether the mode is extended all over the waveguide or localized. To form an $N$-long cavity from a sPhCW, we omit $N$ row of holes in the triangular lattice. In analogy to an LN cavity described in the literature~\cite{okano_analysis_2010}, we denote our system as an sLN cavity. The field confinement in disordered sLN cavities is determined by the local structural disorder and by the localization length $\xi$ compared to the cavity length $L=Na$. When the length of the cavity is sufficiently longer than $\xi$, we can characterize the ballistic-localization crossover. We do that here by calculating the electromagnetic-field quasi-normal-modes (QNMs) of a set of sLN cavities of length $L=40a$. This length is smaller than the lengths employed in our experiments and the simulation domains analyzed in Fig.~\ref{fig:LocsPhCW}, but the full-vectorial three-dimensional calculation of the QNMs is a computationally expensive task, especially with the mesh sizes required to capture the rough slot sidewalls which we model here as in Fig.~\ref{fig:LocsPhCW}. We obtain the group index $n_{\text{g}}$ of the propagating mode from the FSR of the resulting set of QNMs, i.e., $\Delta\nu_{FSR}=\nu_{i+1}-\nu_{i}$, as,
\begin{equation}
\label{eq:ngfromFSR}
n_g(\nu)=\frac{c}{\Delta\nu_{FSR}L}
\end{equation}
where the frequency $\nu$ at which $n_{\text{g}}$ is evaluated is taken as the mean value $\nu=(\nu_i+\nu_{i+1})/2$. The extracted $n_{\text{g}}^{\text{sL40}}$ is plotted (white dots) in Fig.~\ref{fig:sLNloc}a for a perfect ($\sigma=0$) cavity and for four different realisations of disordered cavities with the same disorder level ($\sigma=3$ nm). We compare these calculations with the group index $n^{\text{sPhCW}}_{\text{g}}(\nu)$ obtained directly from the dispersion of the band (solid light-blue line). As plotted in the top panel of Fig.~\ref{fig:sLNloc}a, we reproduce the $n_{\text{g}}$ for the unperturbed sL40 cavity with high accuracy even for very small $N$ (not shown). When the slot line-edge roughness is included in the geometry, the calculated $n^{\text{sLN}}_{\text{g}}(\nu)$ deviates from $n^{\text{sPhCW}}_{\text{g}}(\nu)$ since the frequency of the disordered-induced localized cavities is not determined by the length of the enclosed waveguide, as plotted in the panels 1 to 4 in Fig.~\ref{fig:sLNloc}a. To visualize the random-interference nature of the disorder-induced modes, we plot in Fig.~\ref{fig:sLNloc}b the electromagnetic-field $y$-component corresponding to the first ten QNMs of the structure labelled as "i" in Fig.~\ref{fig:sLNloc}a. Above frequencies corresponding to a theoretical $n_g \geq 15$, the eigenmodes differ from FP cavity modes and are instead strongly localized within the waveguide. This gives rise to fluctuations in their calculated frequencies which lead to strong fluctuations in the extracted $n_{\text{g}}$ using Eq.~\ref{eq:ngfromFSR}. We adopt this to detect the onset of localization in our waveguide as shown in Fig.~2c in the main text. \\

\section{S2. Sample fabrication.}
\label{sec:fab}

The results shown in the main text are obtained from two different sets of samples fabricated using different fabrication processes and commercial silicon-on-insulator (SOI) stacks. In the first set, the device-layer is 240 nm thick undoped silicon and the buried oxide is 3 $\mu$m thermal oxide. The patterns are written using 100 keV electron-beam lithography (JEOL-9500FSZ) on a spin-coated 180 nm chemically semi-amplified electron-sensitive resist (AR-P CSAR6200.09) softmask coated with a 20 nm thermally-evaporated aluminium decharging layer. The mask is transferred into the silicon device-layer using a fluorine-based cryogenic (-19 $C^{\circ}$) ICP-coupled reactive-ion etch passivated using conventional fluorocarbons (C$_{\text{4}}$F$_{\text{8}}$). The structures are released by isotropically etching the buried oxide in buffered hydrofluoric acid (BHF) with a wetting agent, which results in a $\sim$ 3.5 $\mu$m underetched region around the devices. The thicknesses for the device and oxide layers in the second set of samples are 220 nm and 2 $\mu$m, respectively. The fabrication process is more complex than the first.  A multi-layer hardmask consisting of 30 nm  of poly-crystalline chromium followed by  12 nm poly-crystalline silicon is deposited prior to application of a $\sim$ 50 nm  resist (AR-P CSAR6200.09 diluted 1:1 in anisole). After e-beam lithography~\cite{marcus}, the pattern is transferred into the silicon device-layer by successive etch steps using a low-power switched reactive-ion etch, the recent clear-oxidize-remove-etch (CORE) sequence~\cite{CORE,marcus}, modified for the hardmask stack~\cite{CORE2}. Specifically, we etch the poly-crystalline silicon mask in 4 cycles with 20 s etch time and 2 s oxidation, next we strip the resist in a 30 W O$_{\text{2}}$-plasma for 5 min, and etch the poly-crystalline chromium in 65 cycles at 40 W. The device layer is etched in 25 cycles using 20 s etch-time, 18 W platen power in the remove step, and 3 s oxidation. Finally, the remaining chromium mask is stripped by submerging the chip in a commercial solution (Chrome Etch 18) followed by a 2 min dip in a Piranha solution. The structures are released by a 4.0 $\mu$m isotropic underetch of the buried oxide in anhydrous hydrofluoric acid (SPTS uEtch). More details on the fabrication process can be found in Ref.~\cite{marcus}.\\

\begin{figure}[t]
\centering
\includegraphics[width=0.6\textwidth]{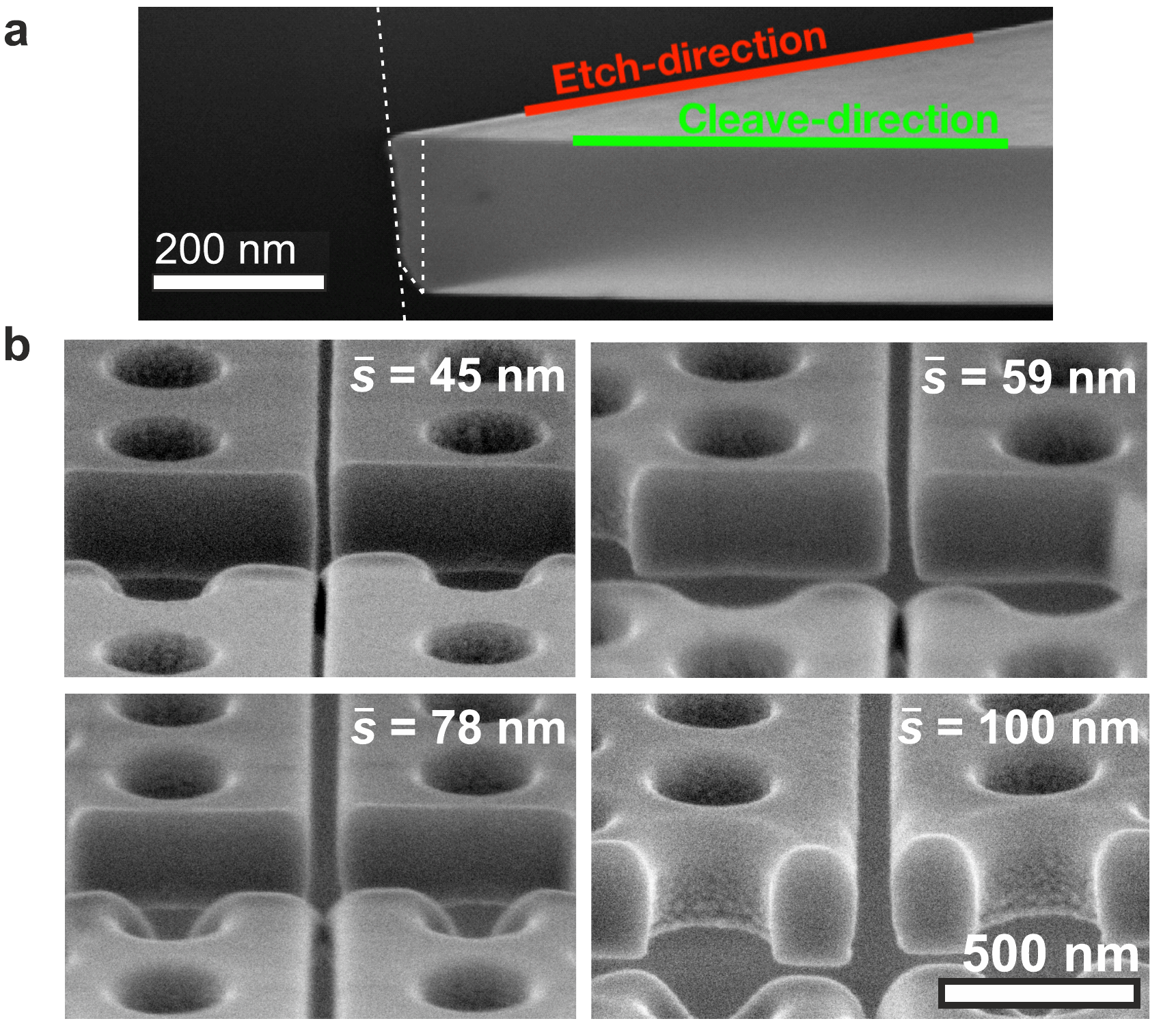}
\caption[Etch verticality in fabricated sPhCWs]{\textbf{Etch verticality in fabricated features.} \textbf{a}, Etch profile along a [1 0 0] cleaved trench, with white dashed lines highlighting a possible geometrical representation of the sidewall. \textbf{b}, Nearly-tangent SEM micrographs of focused-ion beam cut slots, showing the vertical profile of the slot trenches.}
\label{fig:S_vert}
\end{figure}

After fabrication, the devices are thoroughly characterized using scanning electron microscopy (SEM) in order to extract the as-fabricated geometrical parameters, notably the average in-plane parameters, $\{\overline{s},\overline{r},\overline{a}\}$. The post-processed device-layer thickness $t$ and refractive index $n_{\text{Si}}$ are obtained from variable-angle ellipsometric spectroscopy measurements. We estimate the sidewall angle from cross-sectional SEM on [1 0 0] cleaved trenches. While the second set of samples exhibited perfectly vertical sidewalls, the first set did not, as shown in Fig.~\ref{fig:S_vert}a. Both a sidewall angle and notching at the bottom of the layer are observed. Such profile can be described with a two slope profile given by parameters $[h_1,\theta_1]$ = [200 nm, $5^{\circ}$] and $[h_2,\theta_2]$ = [40 nm, $30^{\circ}$] (marked by white dashed lines in Fig.~\ref{fig:S_vert}a). Since notching effects are often highly dependent on the exposed feature~\cite{li_technique_2003}, we assess the etch profile at the level of the sPhCWs by cutting some devices using a focused ion beam (FIB) across both the circular holes and the slot trench. SEM micrographs are shown in Fig.~\ref{fig:S_vert}b. Aside from the lack of slot sidewall verticality, some mask erosion is observed, specially on the holes. Despite the non-trivial differences in sidewall verticality for different features, we use the profile measured on the isolated trench as an etch reference for the first set of samples and use it for all features equally in the simulations used for comparison to experimental data. Note that the angled sidewalls lead to considerable differences in the cut-off wavelength between the two sets even if the parameters used for the photonic crystals have been respectively set to target the same wavelengths. The second set is used in the optical characterization results of Figs. 1 and 2 of the main text, while the optomechanical characterization is carried solely with the second.

\section{S3. Slot-mode: far-field optical characterization.}

A series of photonic integrated circuits including sPhCWs of varying length $L$ have been used to study the propagation losses of sPhCWs using the cutback method and to analyze the vertically scattered far-fields. An example of such circuits is given in Fig.~1 in the main text. Unfortunately, systematic shifts in the cut-off wavelength as a function of $L$ and position within the chip prevent the evaluation of the propagation losses as a function of group index, which could cast light on the localization length $\xi$. However, the scattered far fields give clear spatial evidence of the localized nature of modes close to the cut-off wavelength.

\subsection{S3.1. Far-field imaging.}

\begin{figure}[b!]
\centering
 \includegraphics[width=0.9\textwidth]{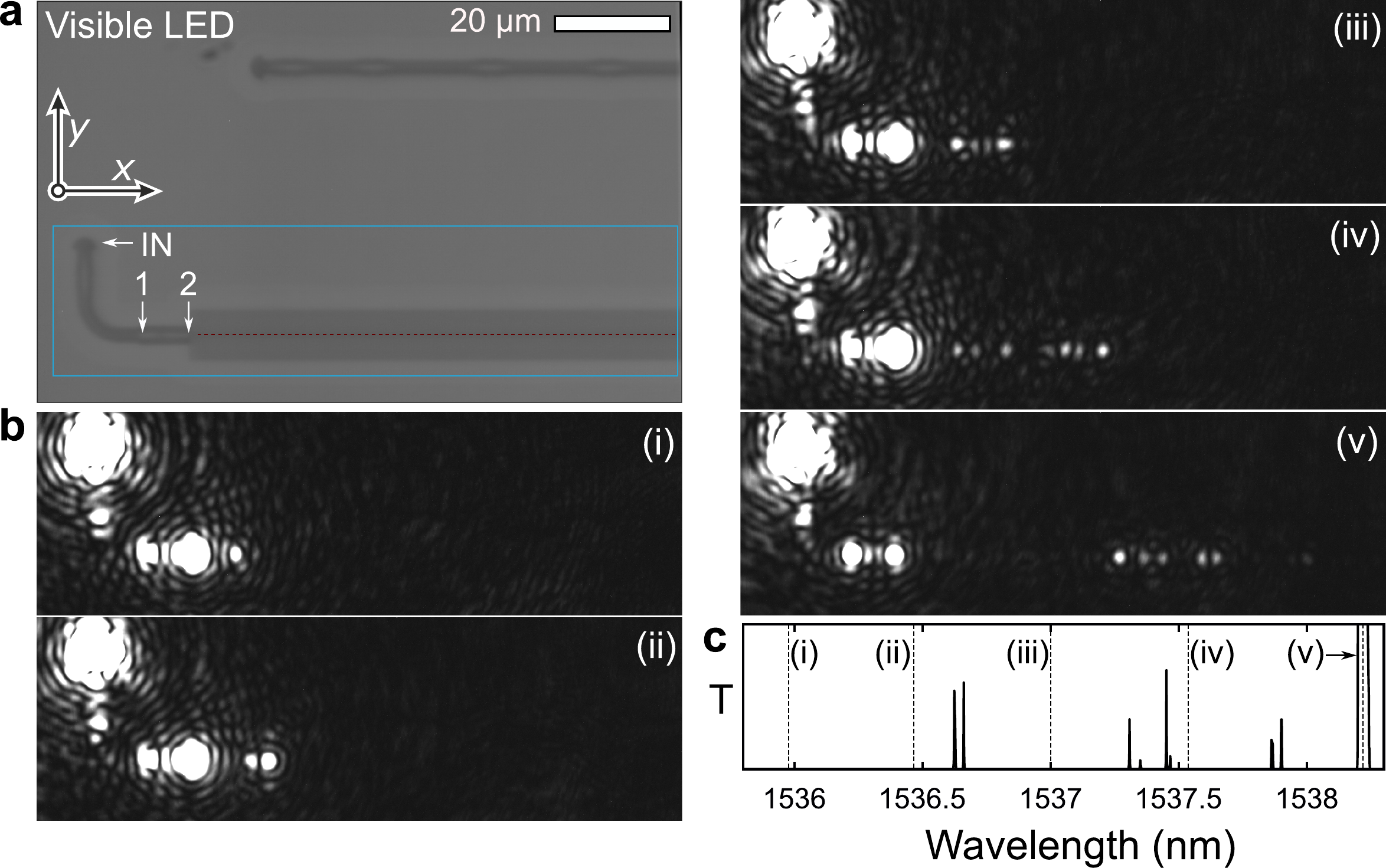}
\caption[Far field emissions]{\textbf{Far field emission from Anderson-localized optical modes.} \textbf{a} Visible image of part of a slot photonic-crystal waveguide circuit ($L$ = 300$a$), including the input and output free-space grating couplers. \textbf{b} Near infrared images of the scattered fields at representative wavelengths. The direct reflection of the input grating coupler leads to the bright spot at the input. The other spot present at all wavelengths corresponds to scattered light reflected at the interface between the slot waveguide and the slot photonic-crystal waveguide. \textbf{c} Transmission measured at the output grating coupler, with the wavelengths shown in \textbf{c} indicated with vertical dashed lines.}
\label{fig:FF}
\end{figure}

We couple light into the circuits by using a long working distance 50X microscope objective (Mitutoyo, NA = 0.42) that focuses light into the grating coupler indicated as "IN" in Fig.~\ref{fig:FF}a. An achromatic half-wave plate is used to excite with polarization parallel to the waveguide axis ($x$), i.e. TE-like. We then collect the vertically scattered fields using the same microscope objective and detect them using a visible and infrared camera (ABA-013VIR). Despite the complex near-field polarization of any Bloch mode, the fundamental slot band (A in Fig.~\ref{fig:sPhCWs}b) has even symmetry relative to the waveguide axis (see Fig.~\ref{fig:sPhCWs}c) and leads to a far-field polarization predominantly perpendicular to the waveguide. This is also expected for Anderson-localized modes resulting from coherent backscattering, which, as described before, can typically be written as the product of an envelope and the Bloch mode. Therefore, we employ cross-polarized excitation-collection by using a linear polarizer aligned to the $y$ axis in the camera path. The spectral map in Fig.~1f of the main text is obtained by step-sweeping a tunable external cavity diode laser (Santec TSL-710, 1480-1640 nm) and acquiring simultaneously the images over a restricted sensor area (indicated by the blue box of Fig.~\ref{fig:FF}a). Afterwards, the signal along the dashed red line is extracted. Fig.~\ref{fig:FF}b shows the images corresponding to the modes highlighted as (i)-(v) in the spectral map of Fig.~1f. The (IN) direct reflection over the grating coupler, the (1) out-of-plane scattering at the strip-to-slot converter and (2) out-of-plane scattering at the interface between the slot waveguide and the sPhcW give rise to intense spots above the saturation level. The rest of the spots result from radiation losses associated to different localized modes. Zooming in into the circuit transmittance shown in Fig. 1f of the main text (Fig.~\ref{fig:FF}c) shows the presence of several small transmission peaks, none of which corresponds to the strongest scattered fields except for mode (v).

\subsection{S3.2. Mode volume estimation.}

\begin{figure}[b!]
\centering
 \includegraphics[width=0.7\textwidth]{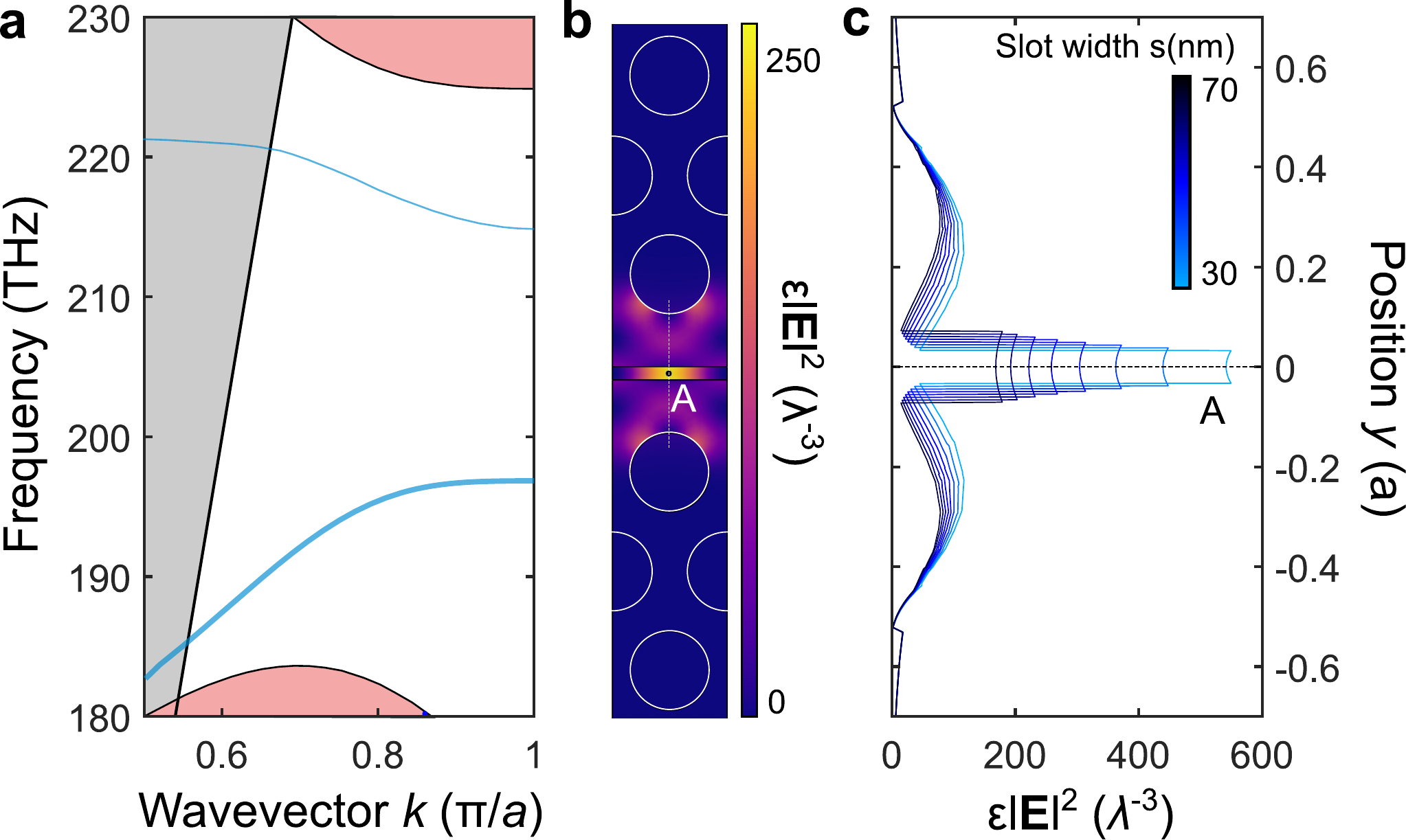}
    \caption{\textbf{Single unit-cell mode volume $V_{\text{eff,cell}}$ in slot photonic-crystal waveguides (sPhcW).} \textbf{a}, Band structure of a sPhcW with pitch $a$ = 450 nm, circle radius $r$ = 155 nm, slot width $s$ = 52 nm and membrane thickness $t$ = 220 nm. The light cone and the bulk bands are shown with shaded grey and red, respectively. \textbf{b}, Normalized electric energy density at the band edge ($k=\frac{\pi}{a}$). \textbf{c}, Normalized electric energy density of the mode of interest along the white dashed line highlighted in \textbf{b}, as a function of the slot width $s$.}
    \label{fig:R1}
\end{figure}

An important property for an optical nanocavity is the level of spatial confinement of the light field. While the precise application targeted for the cavity (Purcell enhancement, two-photon absorption, Kerr non-linearities, etc.) determines the exact figure of merit~\cite{notomi_2010}, a conventional choice in nanophotonics is to use the effective mode volume, $V_{\text{eff}}$. It arises when evaluating the local density of states (LDOS) associated with the coupling of a point-like emitter and an optical cavity. Its canonical definition is
\begin{equation}
    V_{\text{eff}} = \frac{\int_V \varepsilon(\mathbf{r})\left| \mathbf{E}(\mathbf{r}) \right|^2 \mathbf{dr}}{\varepsilon(\mathbf{r_o})\left| \mathbf{E}(\mathbf{r_o})\right|^2}
\end{equation}
where $\textbf{r}_o$ is the position of the emitter. Although this position is not pre-determined in the definition, it is typically taken where the energy density is maximum to define an unequivocal figure of merit. Given the very high quality factors experimentally measured, here we have intentionally avoided taking into account non-hermitian effects~\cite{lalanne_review} into the definition of $V_{\text{eff}}$. From the definition, it is obvious that measuring $V_{\text{eff}}$ would require a near-field measurement to spatially resolve the field. However, some estimate of the expected mode volumes in waveguide-based cavities may be extracted from a combination of far-field experiments and numerical simulations. The main hypothesis for such an estimation is to consider that the localized field can be cast as $\mathbf{E}(\mathbf{r}) \approx f(x)\mathbf{E}_{k}(\mathbf{r})e^{ikx}$ with $f(x)$ an envelope along the waveguide direction ($x$) and $\mathbf{E}_{k}(\mathbf{r})$ the periodic part of the Bloch mode with wavevector $k$. The localized mode also contains the counter-propagating mode at $-k$, but it can be omitted for simplicity. Such an assumption applies well to engineered cavities built from adiabatic potentials~\cite{cavity_envelope} and is also suitable for disorder-induced modes deep in the Anderson-localization regime~\cite{patterson_coherent}. Assuming a slow variation of $f(x)$ relative to the period $a$, the mode volume can be rewritten as
\begin{equation}
    V_{\text{eff}} \approx \frac{\int_x \left|f(x)\right|^2 dx}{a\left|f(x_o)\right|^2}V_{\text{eff,cell}}
    \label{eq:approxV}
\end{equation}
with $V_{\text{eff,cell}}$ the mode volume extracted at the unit cell level. If the far-field intensity is proportional to the field intensity $\left|f(x)\right|^2$, we can infer $V_{\text{eff}}$ directly from the measured far-field and the simulated value of $V_{\text{eff,cell}}$. Thus, the value of $V_{\text{eff,cell}}$ becomes a critical parameter when assessing the order of magnitude of the mode volume of spatially-localized modes in photonic-crystal waveguides.\\

In Figure~\ref{fig:R1} we explore the achieved $V_{\text{eff,cell}}$ for the slot mode of a slot photonic-crystal waveguide, whose dispersion for a slot width $s$ = 52 nm is given in Fig.~\ref{fig:R1}\textbf{a}. The rest of the parameters are selected to be those of the second set of samples. When a Bloch mode is normalized as $\int_V \varepsilon(\mathbf{r})\left| \mathbf{E}_{k}(\mathbf{r}) \right|^2 \mathbf{dr} = 1$, the effective mode volume in the unit cell $V_{\text{eff,cell}}$ is directly evaluated by the inverse of the electric energy density at the relevant point $\textbf{r}_o$. If we chose that point to be at the center of the slot (labeled \textbf{A} in Fig.~\ref{fig:R1}\textbf{b}), where one could imagine having a trapped atom as a quantum emitter, then Fig.~\ref{fig:R1}\textbf{b} evidences that $V_{\text{eff,cell}}\sim \frac{1}{250}\lambda^3$. In addition, this value can go down to $V_{\text{eff,cell}} \sim \frac{1}{550}\lambda^3$ for $s$ = 30 nm, which corresponds to the smallest gaps that are etched through with our current fabrication flow. Using the measured far-fields we can extract the envelope's intensity $\left|f(x)\right|^2$ of the observed modes. In particular, we show in Fig. 1e of the main text a mode with an effective extent of 2.5 $\mu$m, which corresponds to approximately 5 unit cells for the pitch $a$ = 450 nm. Using the approximation in Equation~\ref{eq:approxV}, we estimate a mode volume $V_{\text{eff}}$ as low as  $\left(\frac{2.5 \mu \text{m}}{450 \text{nm}}\right)\frac{1}{250}\lambda^3 \sim \frac{1}{50}(\lambda)^3$.\\

\section{S4. Slot-mode: near-field optical characterization.}
\label{sec:slotmodeoptics}

Butt-coupling from an integrated access waveguide into the slow-light region of a sPhcW, as described in the previous section, is inefficient. Among the wide range of methods to improve the coupling efficiency into sPhCWs~\cite{bogaerts_basic_2004}, we use the evanescent coupling from an optical tapered fiber loop~\cite{michael_optical_2007}, which allows the extraction of dispersion properties from photoni-crystal waveguide resonators.\\

\begin{figure}[b!]
 \centering
  \includegraphics[width=0.65\textwidth]{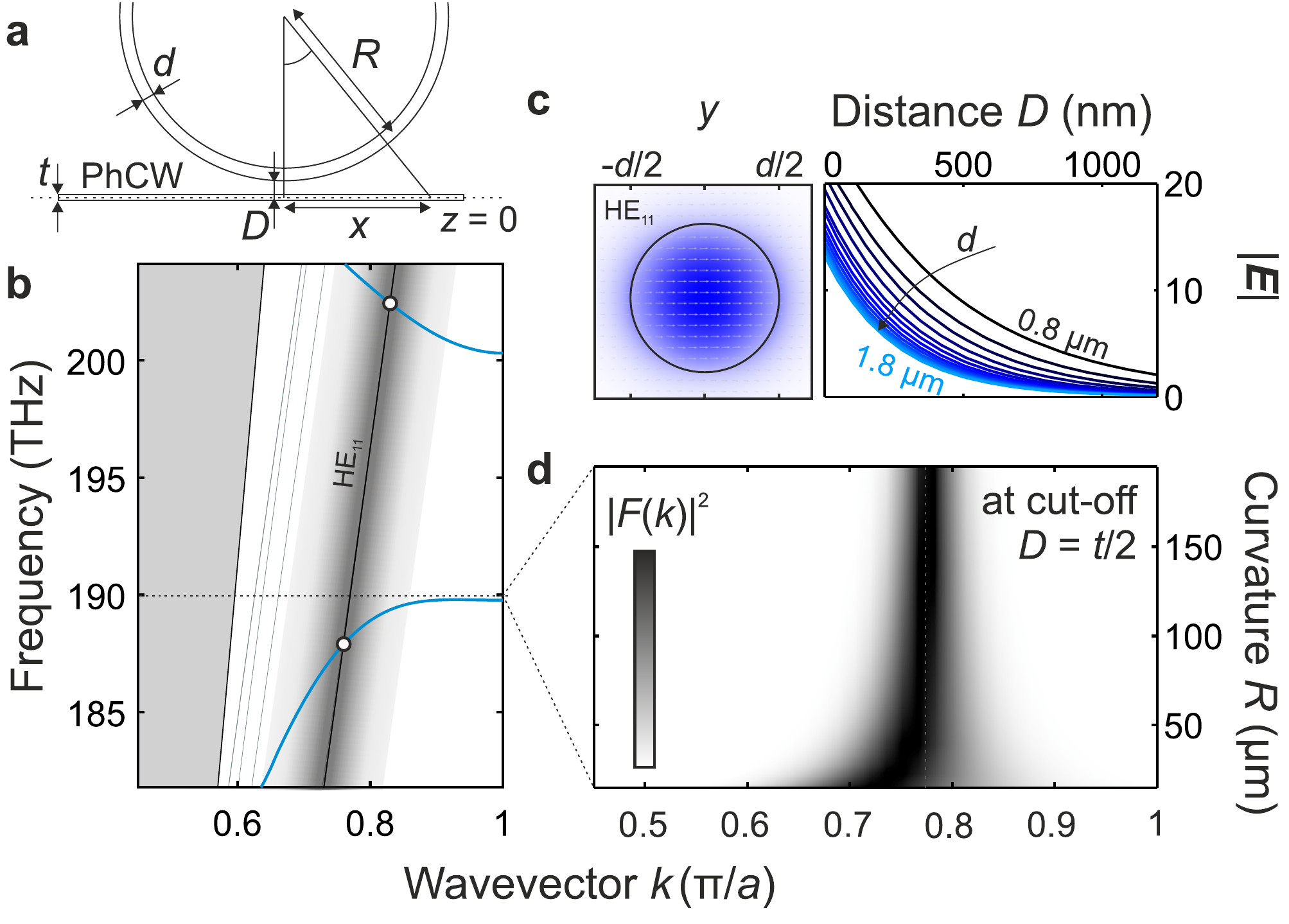}
\caption[Fiber-taper coupling to a photonic crystal waveguide (PhCW)]{\textbf{Phase-matching between an air-cladded fiber and a slotted photonic crystal waveguide (sPhCW).} \textbf{a}, An air-cladded fiber with curvature $R$ and diameter $d$ is placed at a distance $D$ from the mid-plane of a slab of thickness $t$ in which a sPhCW is patterned. \textbf{b}, Band-diagram portrait of the phase-matching condition between the fiber taper and the sPhCW. Waveguide bands are shown in light blue and the modes of a fiber taper of diameter $d$=1.5 $\mu$m are shown in black (HE$_{11}$-mode) and light grey. For a straight fiber taper, the phase matching condition is only achieved at the white-filled dots, while a finite curvature $R$ relaxes the condition by increasing the $k$-space distribution of the fiber taper mode in the PhCW plane $z=0$, as schematically shown by the shading around the HE$_{11}$-mode. \textbf{c}, Electric-field norm, $\lvert \mathbf{E} (x,y)\rvert$, and vector field (arrows) of the fiber HE$_{11}$-mode at the cut-off frequency $f_{\text{co}}$ (left) and its decay in the air cladding (right) for different diameters, $d$. \textbf{d}, Approximated Fourier transform of taper mode at the PhC plane as a function of $R$, showing how the phase-matching region broadens with decreasing $R$. }
\label{fig:kspace}
\end{figure}

\subsection{S4.1. Phase matching considerations.}
\label{subsec:phasematch}

The evanescent coupling between two propagating optical modes occurs whenever they are phase-matched, i.e., when they their dispersion relations intersect. This is indicated by solid white dots in Fig.~\ref{fig:kspace}a, where the dispersion of a sPhCW (solid blue lines) and of an air-cladded fiber taper of diameter $d$ =1.5 $\mu$m (solid black and grey lines) are shown. We consider here the fundamental HE$_{11}$ for the phase-matching condition with fundamental (A) and first-high order (C) odd guided modes (see Fig.~\ref{fig:sPhCWs}c for reference). Since only the intersection points lead to efficient energy transfer~\cite{grillet_efficient_2006}, a full characterization of the band structure of the waveguide would require measurements with fiber-tapers of different (and well-known) diameters~\cite{barclay_probing_2004}. An alternative approach was suggested in Ref.~\cite{lee_characterizing_2008}, where a single taper with a low curvature radius $R$ allows coupling over a much higher bandwidth. This can be understood by mapping the guided electromagnetic field through the curved fiber to the PhC plane and taking its Fourier transform. This last approach can be approximated as
\begin{equation}
\label{eq:FTcurvedfiber}
F(k_z)=\int_{-\infty}^{\infty}f(0,\sqrt{x^2+(R+D)^2}-R)e^{i\beta R \text{ atan}(\frac{x}{R+D})}e^{ik_zz}dz
\end{equation}
where the curvature is considered small enough to maintain the transverse profile $f(x',y')$ of the straight fiber at any $\theta$, with $\theta$ and all other parameters defined as shown in the upper parf of Fig.~\ref{fig:kspace}. Such quantity is shown in Fig.~\ref{fig:kspace}d and shows how a smaller radius $R$ leads to a a reciprocal space broadening which facilitates the phase-matching condition over a large portion of the dispersion. The diameter of the fiber taper also plays a very important role. A larger diameter taper reduces the HE$_{11}$ slope shown in Fig.~\ref{fig:kspace}a, which is beneficial to couple light into the slow-light region of the slot-guided mode of our sPhCW. However, this also leads to a thinner evanescent field extension, as seen in the right panel of Fig.~\ref{fig:kspace}c, which considerably reduces the coupling for a fixed distance $d$. In addition, a larger diameter facilitates the presence of other higher order propagating modes which can lead to beats between different propagating modes in the taper when driven from either side. In our case, the fiber diameter of 1.5 $\mu$m simulated here is selected to allow a sufficient exponential tail at $D$ = 120-125 nm, i.e., half the thickness of the used slabs, while maintaining the center of the broadened dispersion close to the cut-off of the $y$-even slot-guided band.\\

\begin{figure}[b!]
\centering
\includegraphics[width=0.7\textwidth]{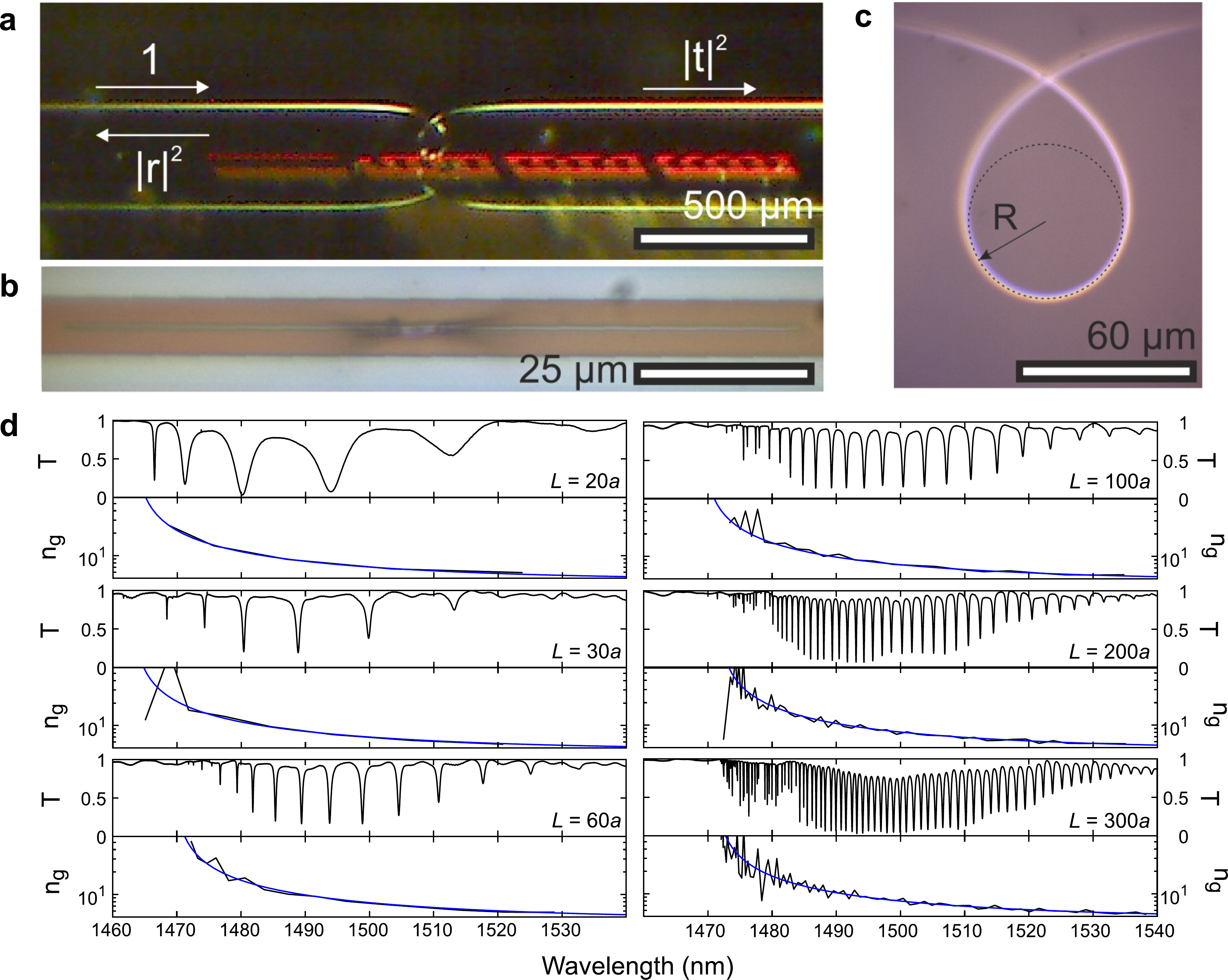}
\caption[Fiber-taper evanescent coupling experimental setup]{\textbf{Evanescent coupling to Anderson-localized and Fabry-Perot modes in a silicon slot photonic-crystal waveguide.} \textbf{a}, Lateral image of the employed optical excitation scheme. \textbf{b}, Microscope image of the loop, with the curvature $R$ schematically shown. \textbf{c}, Microscope image of the sPhCW surface with the loop in contact and aligned with the waveguide axis. \textbf{d}, Normalized transmission for sLN cavities ($\overline{s}$ = 70 nm) of different length $L$ and comparison of the simulated and extracted group indices. The geometry used in simulation is obtained as average parameters from a top-view SEM and setting $t$ = 220 nm and $n_{\text{Si}}=3.48$. The sidewalls are considered to be vertical.}
\label{fig:sLN_exp}
\end{figure}

\subsection{S4.2. Experimental setup.}
\label{subsec:setup}

We produce an adiabatic fiber taper by stretching a telecom single-mode optical fiber in a controlled way via two independent motorized stages while heating the central part to a temperature of 1180$^{\circ}$C using an electric ceramic micro-heater~\cite{ding_ultralow_2010}. The micro-loop is created by a twist-pull-push process and in addition to broadening the phase-matching condition, allows spatially local probing. A lateral image of the resulting fiber loop over a photonic chip is shown in the setup schematic of Fig.~\ref{fig:sLN_exp}a. For the optical characterization, a tunable near-infrared (NIR) external-cavity diode laser (Yenista T100S-HP) is used. A variable optical attenuator (VOA) and a fiber-polarization controller (FPC) are included to control the power and polarization of the optical excitation. The one-sided optical driving configuration is sketched with arrows, allowing transmission, $T=\lvert t \rvert^2$, and reflection, $R=\lvert r\rvert^2$, measurements by using an optical circulator. The loop is manually placed along the axis of the waveguide, as shown in the microscope image of Fig.~\ref{fig:sLN_exp}b, and set to be at the minimum practical radius $R$ = 50 $\mu$m, as shown in Fig.~\ref{fig:sLN_exp}c. Light coupled out from the waveguide is detected with a wavelength-calibrated multiport power-meter (Agilent 8164B) for optical spectroscopy and/or with a fast 12 GHz photoreceiver (New Focus 1544-B) for mechanical spectroscopy. In the latter case, the signal can be simultaneously read out with a 4 GHz digital oscilloscope (Tektronix TDS7404) or a 13 GHz electronic spectrum analyzer (Anritsu MS2830A). The experimental setup also includes an electro-optic phase modulator (iXblue Photonics MPZ-LN-10, $V_{\pi}$ = 4.9 V) driven by a vector network analyzer (Rohde\&Schwarz ZVA 50), which are used for optomechanical characterization (see Section S6).\\

\subsection{S4.3. Group index extraction.}
\label{subsec:ngextraction}

We apply the method to determine the crossover from ballistic propagation to localization by measuring the optical transmission of a set of sLN cavities of length $L$ = [20, 30, 60, 100, 200, 300]$a$, with the fiber loop placed at the center of the cavity. The top panels in Fig.~\ref{fig:sLN_exp}d plot characteristic optical transmission spectra for waveguides with the different length. Using Eq.~\ref{eq:ngfromFSR}, we extract the group index of the mode from the FSR in each spectrum, plotted as black solid curves in Fig.~\ref{fig:sLN_exp}. The measurement of the group index as a function of length is a key control experiment to confirm the onset of Anderson localization~\cite{chirag}. The simulated group index (blue solid curve) has been wavelength-shifted to produce the best fit between experiment and theory over the region where clear Fabry-P\'erot resonances are visible. The onset of group index fluctuations qualitatively shifts towards larger values of $n_{\text{g}}$ for shorter cavities. For very short waveguides, e.g. $L = 20a$, we do not observe any significant fluctuation in the transmitted intensity and the group index can be smoothly reproduced as a Fabry-P\'erot cavity. However, for longer cavities this does not hold anymore. For $L = 100a$, we already observe significant fluctuations in the transmitted light and in the group index estimated from the measurements. This is a clear fingerprint of the dispersive nature of the localization length and the dependence of the occurrence of localization on the length of the cavity.

\subsection{S4.4. Transmission fluctuations and dimensionless conductance.}
\label{subsec:fluctuations}

\begin{figure}[b!]
\centering
 \includegraphics[width=0.7\textwidth]{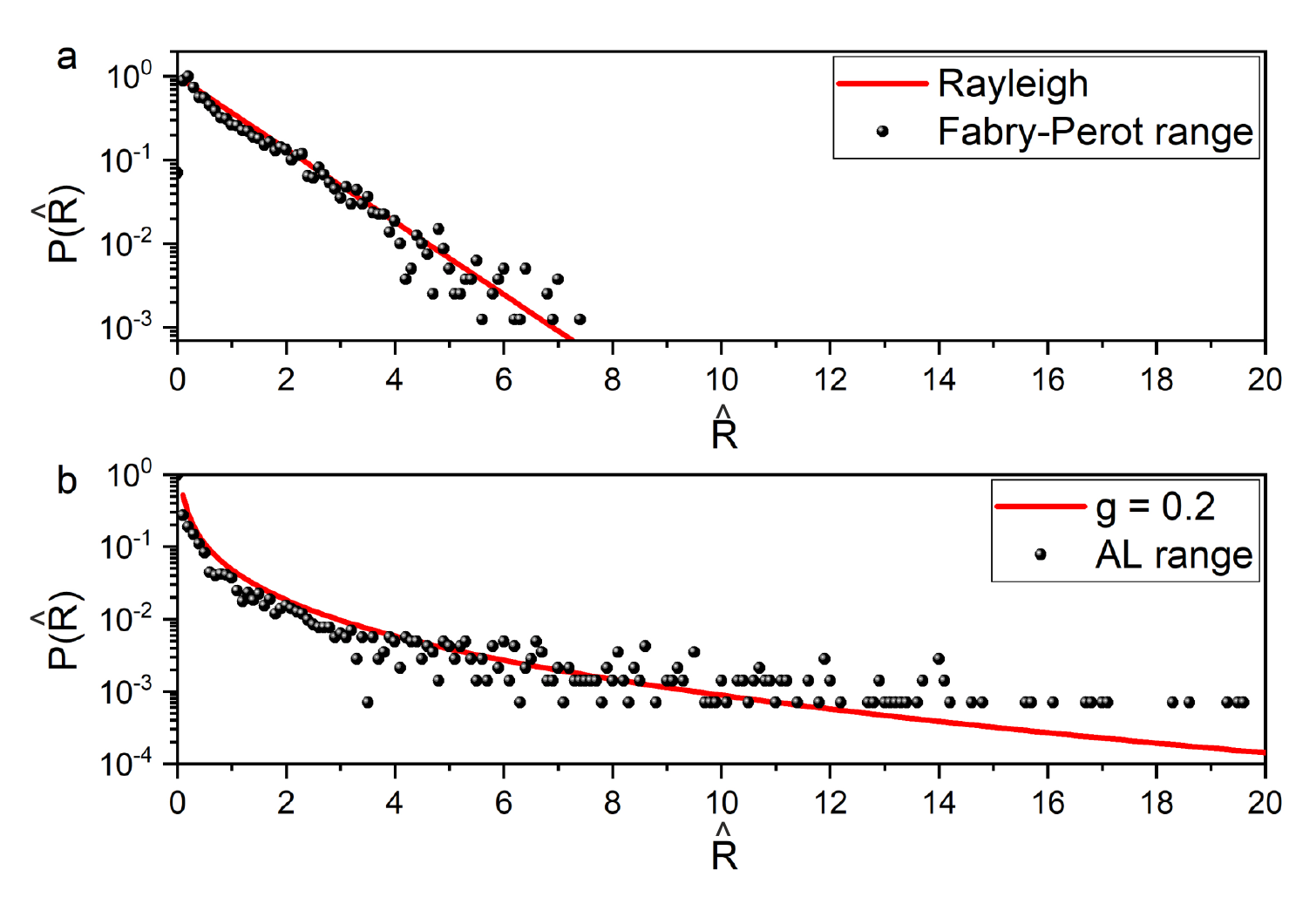}
    \caption{\textbf{Statistical fingerprints of Anderson-localization of light in slot photonic-crystal waveguides.} Normalized intensity distribution recorded from the transmitted signal through the tapered fiber in evanescent coupling with the waveguide. While \textbf{a} limits the distribution analysis to the spectral range where Fabry-P\'erot (FP) are observed, \textbf{b} quantifies the distribution in the Anderson localization (AL) spectral range. We discriminate the FP range from the AL one when the extracted group index from the spectra departs significantly from the calculated one (see Fig. 1\textbf{e} in the original manuscript).}
    \label{fig:R3}
\end{figure}

The use of the variance in the transmitted or reflected intensity has been routinely used as a proof of Anderson localization (see Ref.~\cite{Annalen}, section 3 for a summary). The fluctuations of the intensity of the transmitted light are analyzed instead of the average transmission itself. This approach helps discriminating multiple scattering from material absorption~\cite{Chabanov}. It is a very useful and practical approach as it only requires collecting many optical transmission (or reflection) spectra. Once measured, these spectra are normalized by dividing the raw data by the mean value of the transmission/reflection intensity. Finally, the histogram of the normalized intensity is obtained, see Figure~\ref{fig:R3}, revealing a lot information about the multiple scattering regime. Deep in the localization regime, this histogram leads to heavy-tailed log-normal distribution which can be interpreted as follows: the average transmission/reflection is very low but the probability of very intense dips/peaks is not negligible. Such a log-normal distribution is characteristic of the localization regime and it has been used in electronic, photonic and phononic systems to characterize the dimensionless conductance of the system~\cite{Rossum}. In the past, we have also used this approach in a \textit{shamrock} photonic-crystal waveguide~\cite{PRB_SP} where the measurements were carried out using the same type of coupling. The analysis of the variance in the Anderson-localized region and the Fabry-P\'erot region of the waveguide is shown in Figure~\ref{fig:R3}. For a moderate group index, the role of disorder in negligible and Fabry-P\'erot resonances can be measured which leads to a transmission intensity well described by Rayleigh statistics (Fig.~\ref{fig:R3}\textbf{b}). For larger group index, however, the effect of disorder is significant and $P(\hat{R})$ evolves into a heavy-tailed distribution, revealing the presence of few but very bright peaks in a low-intensity background. The pronounced log-normal distribution intensity extracted from the AL spectra is a clear hint that the system is deep in the localization regime within this spectral range from which we extract a dimensionless conductance of $g = 0.2$ as the only fitting parameter to the M. C. W. van Rossum and Th. M. Nieuwenhuizen theory~\cite{Rossum,Annalen}.\\

\subsection{S4.5. Quality factor distributions.}
\label{subsec:Qfactors}

\begin{figure}[b!]
\centering
 \includegraphics[width=0.7\textwidth]{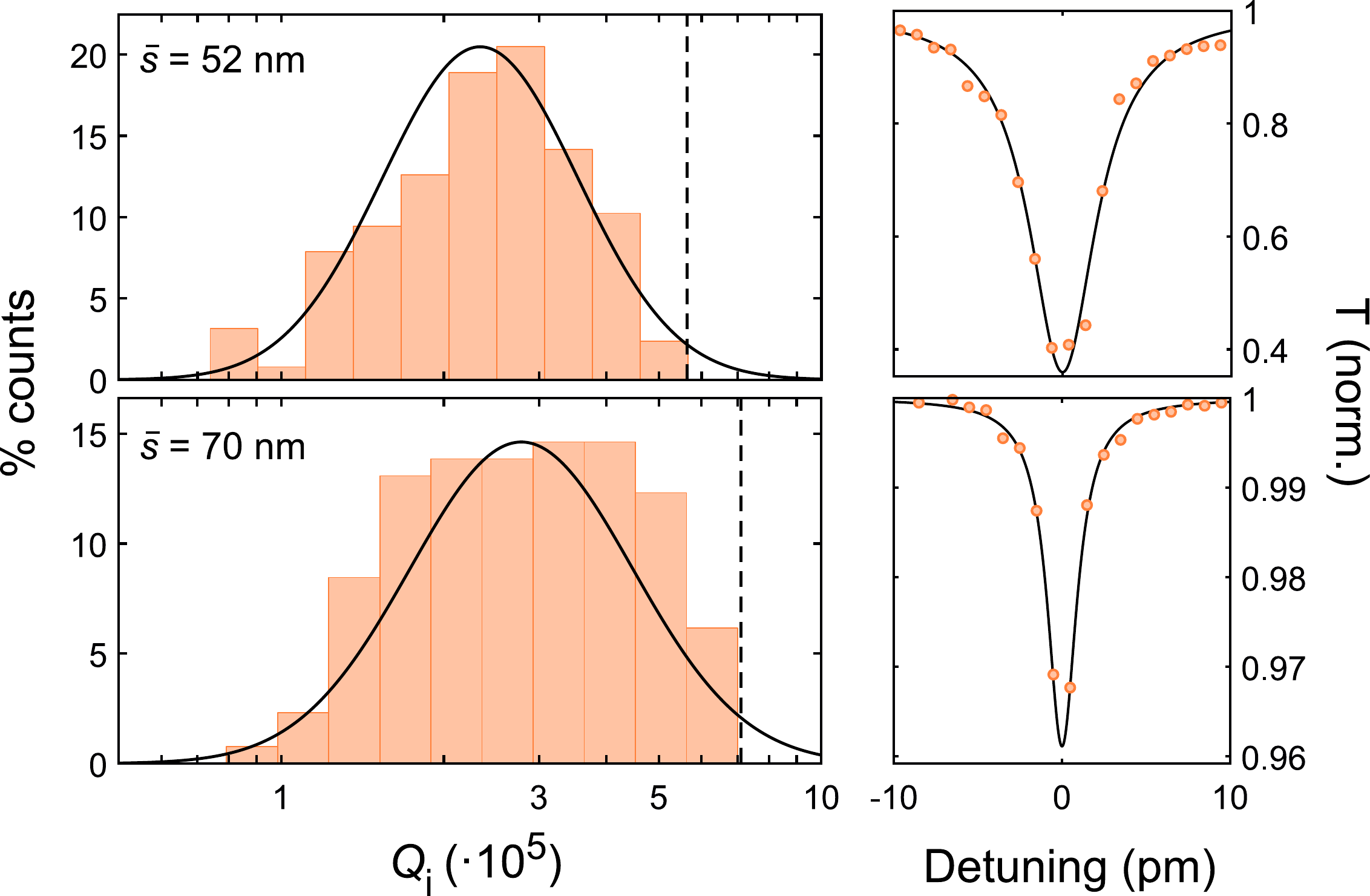}
\caption[Quality factors]{\textbf{Quality-factor distribution of Anderson-localized modes in a slotted photonic-crystal waveguide} Histograms of $Q_{\text{i}}$ for the Anderson-localized modes measured at the center of 11 nominally identical sL300 cavities for slot widths (top) $\overline{s}$ = 52 nm and (bottom) 70 nm. Dashed lines indicate the highest $Q_{\text{i}}$ measured for both slot widths, the optical transmission across which is also given.}
\label{fig:Qsnew}
\end{figure}

From the transmission spectra shown in Fig.~\ref{fig:sLN_exp}d, we can extract the quality factor of the Anderson-localized optical modes, which depend on a considerable amount of parameters such as the waveguide length $L$, the slot width $\overline{s}$ and the slot sidewall roughness $\delta s$. We obtain the statistical distribution by measuring (at least) eleven different optical spectra for a given pair $\{\overline{s},L\}$. Only the peaks well within the region with high $n_{\text{g}}$ fluctuations is considered, which attests their disorder-induced nature. We compute their intrinsic quality factor, $Q_{\text{i}}$, assuming an intrinsic loss rate, $\kappa_{\text{i}}$, and an extrinsic coupling rate, $\kappa_{\text{e}}/2$, corresponding to forward and backward coupling to the fiber loop~\cite{CMT}. The transmission spectrum corresponding to each mode then has a Lorentzian lineshape with a width given by $\kappa=\kappa_{\text{i}}+\kappa_{\text{e}}$ and an on-resonance transmission dip $T_{\text{r}}$ given by $T_{\text{r}}=(1-\frac{\kappa_{\text{e}}}{\kappa})^2$. We use this shape to fit all measured peaks with either single- or multi-Lorentzians depending on the amount of spectral overlap. $Q_{\text{i}}$ is then obtained from the resonant frequency/wavelength and $\kappa_{\text{i}}$. To quantify the accuracy of the fit, a root-mean-squared (r.m.s.) residual value, $\chi$, is calculated for each fit as:
\begin{equation}
    \chi = \sqrt{\frac{1}{N}\sum_{i=1}^{N}\left(T_{\text{fit}}-T_{\text{experiment}}\right)^2}
\end{equation}
with $N$ the number of sampling points. Some of the observed modes, specially in the first set of samples, exhibited lineshapes closer to a Fano lineshape, resulting in $\chi$ values much larger than the rest of modes and were consequently discarded from the final distribution. Some localized modes appear at different positions within the waveguide. To avoid the double counting them in the statistics, we disregard the modes that present values of $Q_{\text{i}}$ less than 5000 apart and wavelengths than 0.2 nm apart in different spectra. These filtering values are chosen to match the typically observed deviations in wavelength and losses induced by the fiber taper itself acting as a perturbation~\cite{cognee_mapping_2019}. The $Q_{\text{i}}$ of the remaining modes are binned into logspace equidistant bins. The histograms of $Q_{\text{i}}$ are then fit to a log-normal distribution, as observed in state-of-the art conventional photonic-crystal waveguide deep in the localization regime~\cite{vasco_statistics_2017,Smolka}.\\

\begin{figure}[b!]
\centering
\includegraphics[width=0.6\textwidth]{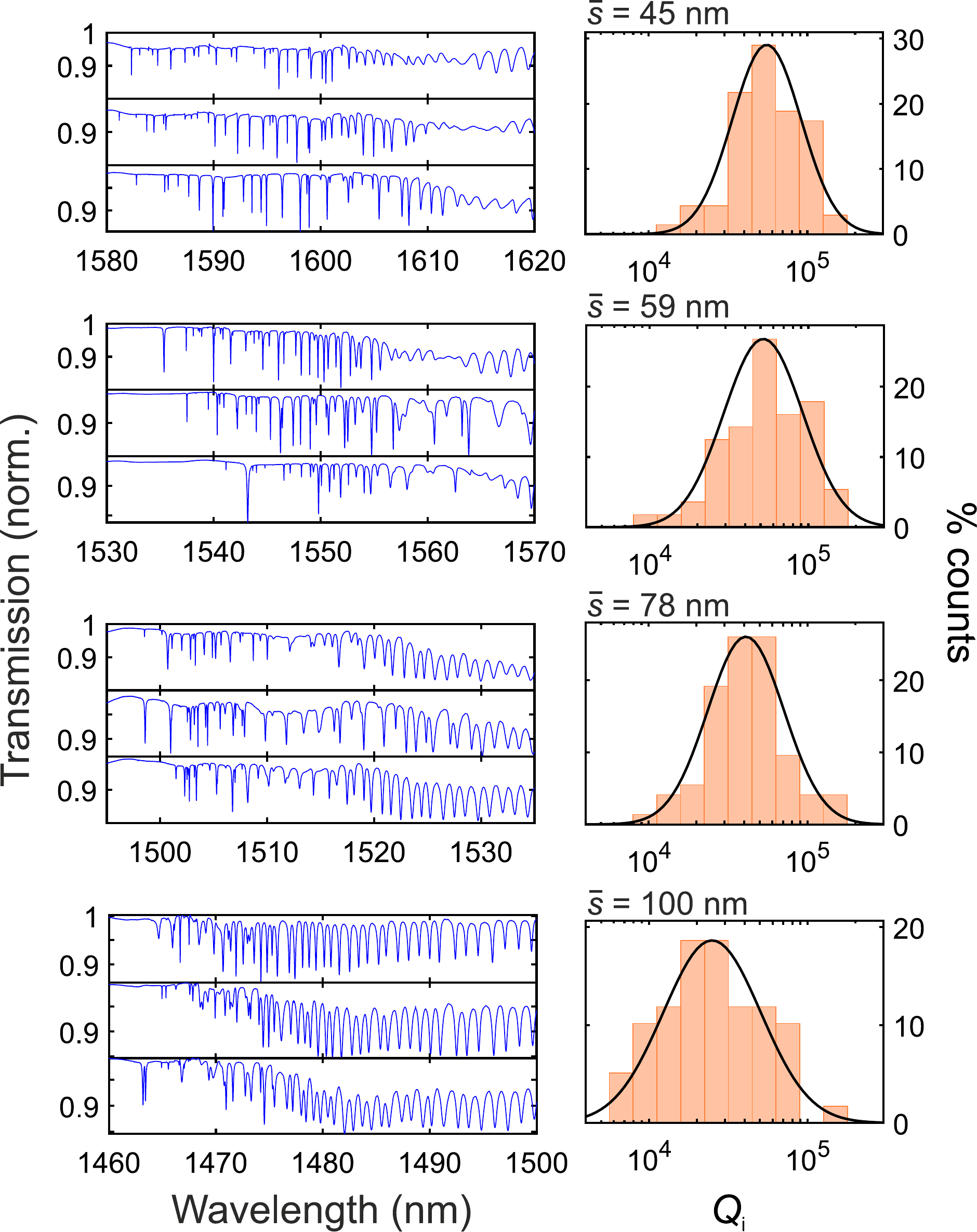}
\caption[radiationpressure]{\textbf{Quality-factor distribution of Anderson-localized modes in a slotted photonic-crystal waveguide with angled sidewalls.} (Left) Three representative optical spectra for each slot size showing both the Anderson-localized region as well as part of the region with Fabry-P\'{e}rot cavity resonances. (Right) Histogram of the log$_{10}(Q_i)$ and the respective log-normal fit (solid orange line).}
\label{fig:Qfacdistribution}
\end{figure}

Fig.~\ref{fig:Qsnew} shows the quality factor histograms and fits for slot widths $\overline{s}$ = 52 nm and $\overline{s}$ = 70 nm in waveguides of length $L$ = 300$a$. The measured distribution is approximately log-normal and practically all quality factors are above $Q_{\text{i}}>10^5$, a \textit{lower bound} also found in shorter waveguides. The distribution moves to higher values for wider slots, which we attribute to the lower roughness and consequent reduction of out-of-plane radiation losses. The dashed lines in the histograms highlight the maximum values measured, the fit to which is also given for reference. These are extracted to be $Q_{\text{i}} = 5.44\cdot 10^5$ for $\overline{s}$ = 52 nm and $Q_{\text{i}} = 7.09\cdot 10^5$ for $\overline{s}$ = 70 nm, both values well above previously reported $Q$s for Anderson-localized modes. Comparison of the histograms obtained for the first and second set of samples described in Section S2 shows that a key parameter in determining the overall distribution and its dependence with $\overline{s}$ is the sidewall verticality. Fig.~\ref{fig:Qfacdistribution} shows relevant optical spectra and distributions for 4 different slot widths $\overline{s}$ in the first sample, which has angled sidewalls as described previously. This breaks the vertical symmetry and promotes scattering of the propagating light (and the localized modes) into a transverse-magnetic-like (TM-like) continuum of delocalized modes, which limits the $Q$ to values around $10^5$, i.e., it lowers the $Q$s by one order of magnitude. Contrary to the behaviour shown in Fig.~\ref{fig:Qsnew}, such scattering mechanism leads to a decrease in $Q$ for wider slots. This occurs because the crossing between TE and TM-like bands moves closer to the band edge (where localization occurs) for wider slots, occurring practically at the band edge for $\overline{s}$ = 100 nm.\\

\begin{figure}[b!]
 \centering
  \includegraphics[width=0.6\columnwidth]{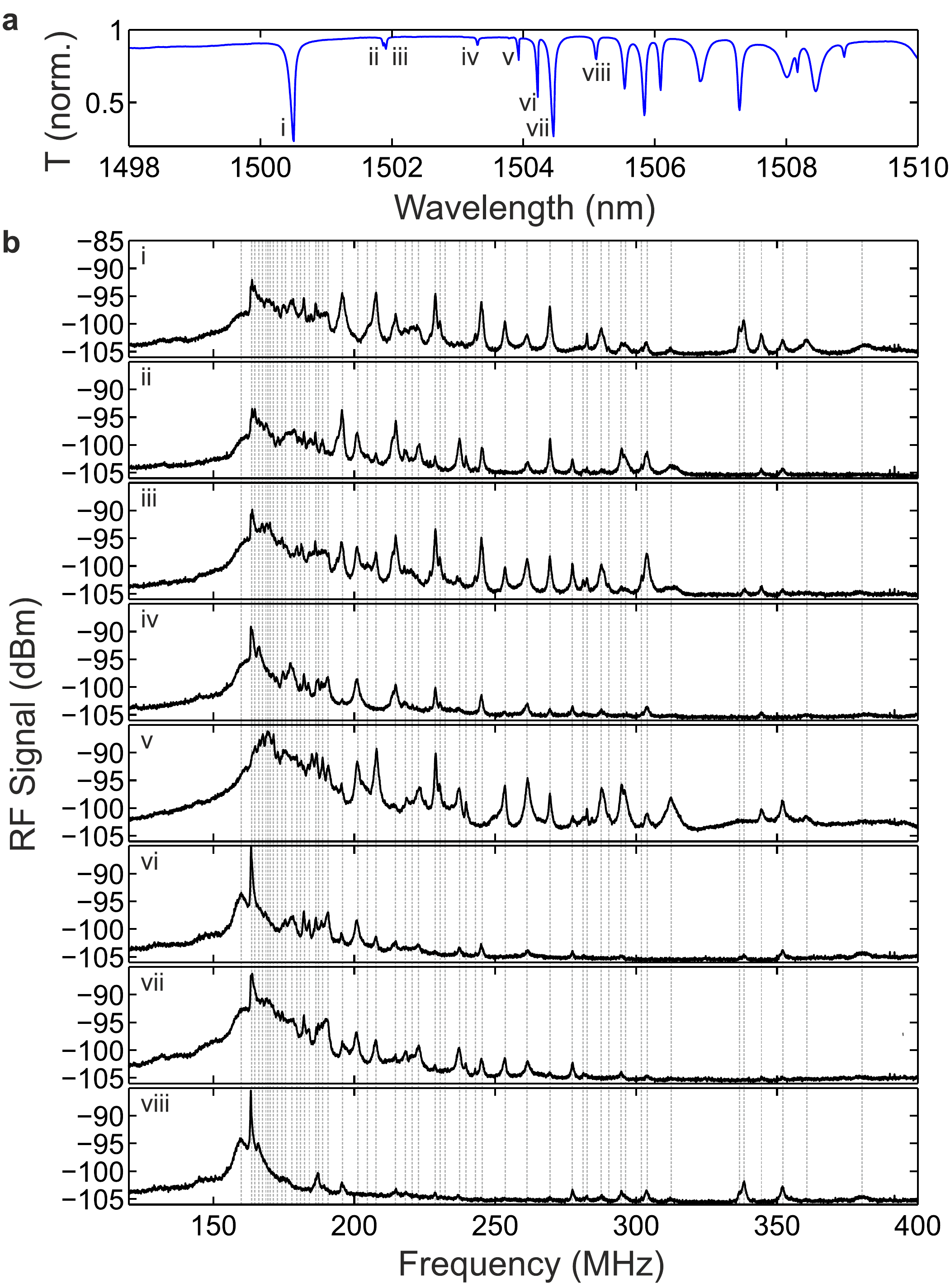}
\caption[Mechanical spectroscopy of a sPhCW]{\textbf{Low-frequency mechanical spectroscopy of a slotted photonic crystal waveguide using multiple optical modes.}  \textbf{a}, Optical spectrum of a $\overline{s}$ = 78 nm cavity-waveguide of length $L=309a$, with $a$ = 470 nm, in the frequency region where light localization occurs. Some of the optical modes are labeled i to viii. \textbf{b}, Radio-frequency (RF) spectra of the transmitted light when probing the modes i-viii, showing multiple Lorentzian peaks in the range from 120 MHz to 400 MHz. The peaks are manually inspected and the dashed lines represent those present in at least two of the RF spectra from a large set of spectra.}
\label{fig:slot78OM}
\end{figure}

\section{S5. Mechanical spectroscopy.}
\label{sec:mecmodes}

We characterize the thermally-activated mechanical motion of the system by measuring the radio-frequency modulation of the Anderson-localized cavities. Fig.~\ref{fig:slot78OM}a plots a low-power optical transmission spectrum of a cavity-waveguide with a number of these Anderson-localized cavity modes labeled as (i)-(viii) and Fig.~\ref{fig:slot78OM}b plots their radio-frequency modulation where a high-drive power of $P_{in}$ = 0.7 mW has been used. This power allows the optical transduction of as many mechanical modes as possible above the noise level without entering into a region with any self-sustained dynamics. The dashed-grey lines across all the spectra mark the central frequencies of the mechanical modes found, whose associated transduction amplitude depends strongly of the optical mode being probed.

\begin{figure}[b!]
 \centering
  \includegraphics[width=0.7\columnwidth]{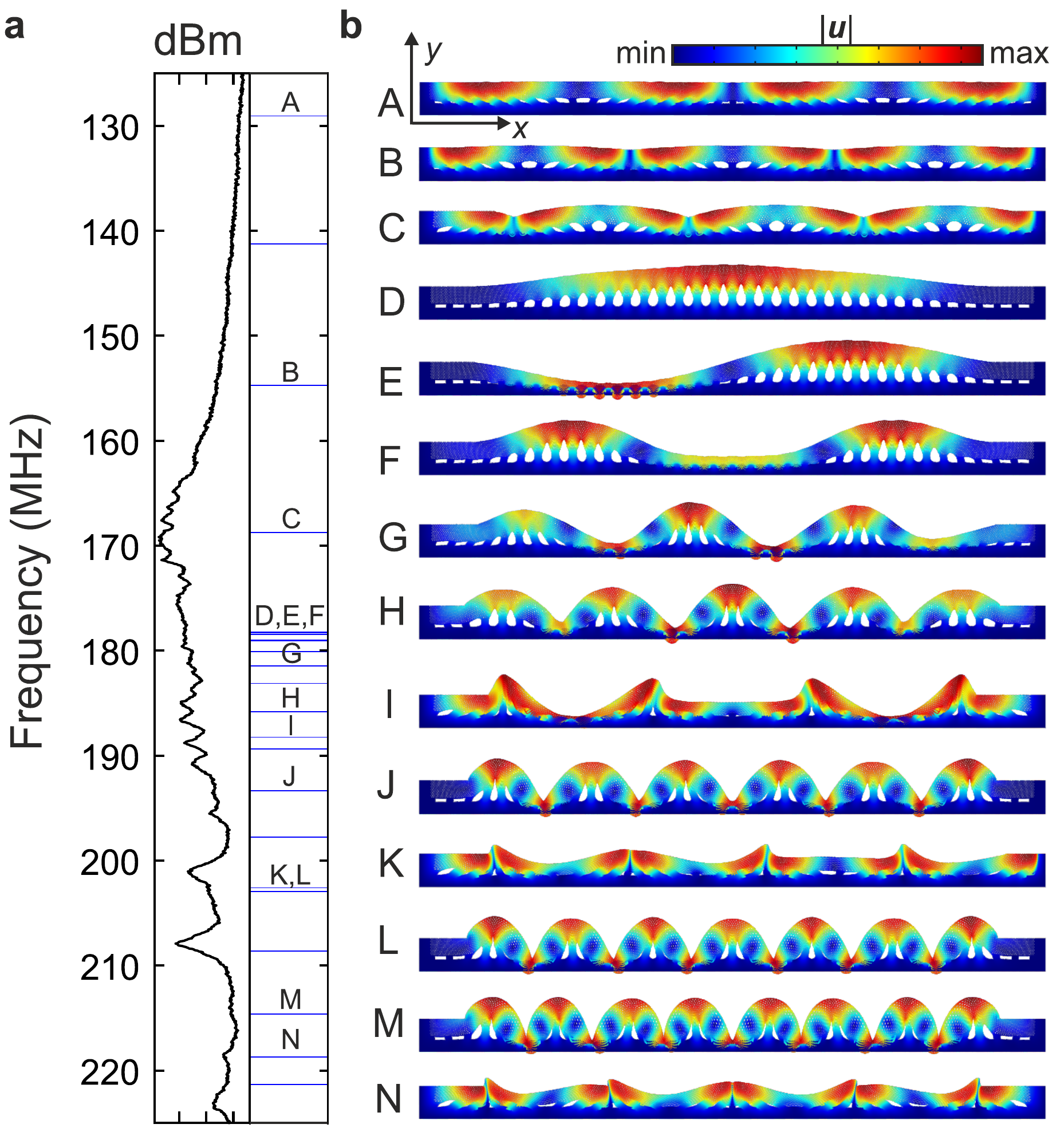}
\caption[Mechanical eigenmodes of the fabricated sPhCWs]{\textbf{Simulated low-frequency mechanical eigenspectrum of a fabricated device.} \textbf{a}, The experimentally extracted eigenspectrum is compared to \textbf{b}, finite-element calculations of the $z$ and $y$-symmetric mechanical modes. Specific modes in the eigenspectrum are labelled from A to N and their deformation profiles given, with the deformation being exaggerated for visualization. }
\label{fig:mechanicalDOS}
\end{figure}

We compare these frequencies with the mechanical eigenspectrum calculated with a finite-element method for this waveguide, as shown in detail in Fig.~\ref{fig:mechanicalDOS}. For this calculation, we consider the exact in-plane geometry of the cavity-waveguide (except for disorder), including the underetched lateral trenches in the SOI wafer with lengths of 3.25 $\mu$m. We impose a fixed boundary condition at the edge of the simulation domain mimicking a non-moving solid. Due to symmetry considerations entering the optomechanical coupling calculations, only modes that are simultaneously $z$-symmetric (mainly in-plane) and $y$-symmetric (both sides of the slot oscillating in anti-phase) are calculated, which reduces the simulation domain to a quarter of the structure. We can classify the modes found in our calculation as a function of the relative contribution of shear (displacements along $x$) or normal displacements (displacements along $y$). The first three modes (labeled A, B and C) are predominantly of shear type and populate the spectrum up to 169 MHz, as shown in Fig.~\ref{fig:mechanicalDOS} where the mode profile  deformation is depicted in addition to the color-scale giving $\lvert u(x,y,z)\rvert$ at the top silicon surface. Another family of modes (labeled D, E, F, G, H, J, L, M), are breathing-like modes with a pronounced mechanical displacement in the $y$ direction, producing an important relative change of the distance between the two slabs. We expect the fundamental in-plane breathing mode (D) to be the predominant peak in all measured spectra and to be the first optomechanically active mechanical mode. Simulations predict a frequency $\Omega_m/2\pi$ = 178 MHz, which is 10 MHz higher than the first mechanical resonance measured. Above this frequency, both the calculated and measured spectra are densely packed with mechanical modes and the overall spectral agreement between them is good. We attribute the overall red-shift and the higher density of modes in experiment to the fabrication imperfection and the vertical profile of the sample, shown in Fig.~\ref{fig:S_vert}a. For example, disorder or the presence of the fiber loop enables the coupling of $y$-symmetric and anti-symmetric mechanical modes leading to two peaks instead of one, while the coupling of in-plane $z$-symmetric and flexural $z$-anti-symmetric modes has a more unpredictable effect. To analyze this spectral mixing in detail, we have calculated the mechanical eigenspectrum of a disordered structure of length $L=100a$ considering both the vertical and angled sidewalls. When the $z$-symmetry is preserved, the $z$-antisymmetric flexural modes densely populate the full frequency range considered here but are, in principle, optomechanically dark. However, some of them are only a few MHz detuned from the main in-plane modes, leading to spectral mixing with in-plane motion when the symmetry is broken. This explains the highly packed spectrum in the 160-180 MHz region, since originally flexural modes become optomechanically active. Additionally, breaking the vertical symmetry leads to a general red-shit of the mechanical spectrum in agreement with our measurement. The high level of transduction of the very low frequency (2-10 MHz) drum modes of the full membrane (see Fig.4d in the main text) also provides an important signature of the role played by the vertical profile in determining the final mechanical eigenspectrum of the structure under investigation. Similar observations have been reported for weak stress-induced bowing in an engineered heterostructure cavity~\cite{safavi-naeini_optomechanics_2010}.\\

We have made the same analysis of Fig. \ref{fig:slot78OM} for the set of sPhCWs with different $\overline{s}$. The type of RF spectra observed, including the precise frequencies of the detected peaks, coincides across all devices since the value of the air-slot width $\overline{s}$ has negligible effect on the MHz modes being probed. However, we measure a stronger transduced signal for narrower air slots which is explained by the higher vacuum optomechanical coupling rate $g_o$ when reducing $\overline{s}$ (see Fig. 3 in the main text).

\section{S5. Optomechanical coupling.}
\label{omcharac}

\begin{figure}
 \centering
  \includegraphics[width=0.6\columnwidth]{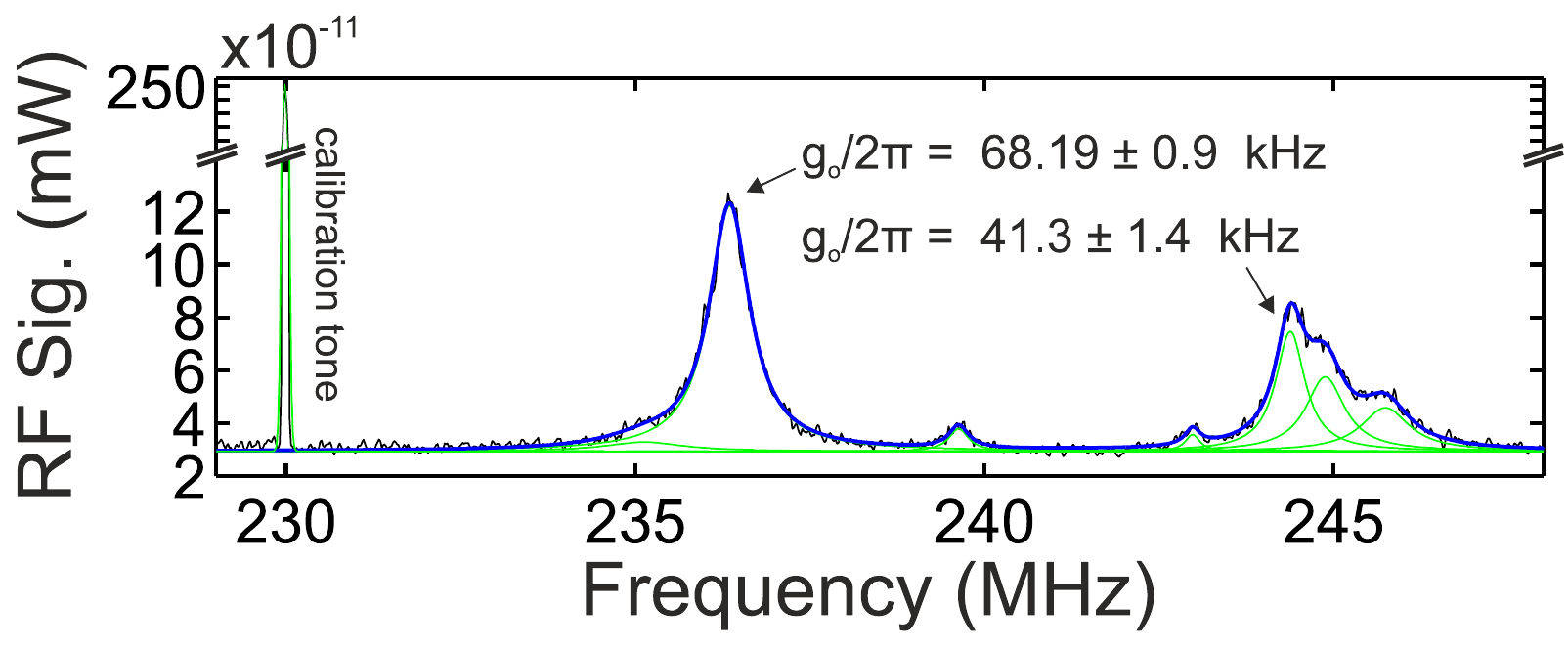}
\caption[Statistics of the measured $g_o$ in disordered sPhCWs]{\textbf{Calibration of the vacuum optomechanical coupling rate, $g_{\text{o}}$, in the Anderson-localization regime.} A frequency/phase-modulation technique is employed to extract $g_{\text{o}}$ between the observed mechanical modes and the driven optical cavity. The mechanical modes are fitted (blue line) using a sum of Lorentzian lineshapes and a background (green lines), while the EOM-generated calibration tone is fitted with a Gaussian lineshape (also green). This procedure is repeated for multiple frequency windows and with multiple localized modes, leading to the statistical distributions shown in Fig. 2 in the main text, where a shift towards higher $g_{\text{o}}$ is seen with decreasing slot width $\overline{s}$.}
\label{fig:OMCouplingExperimental}
\end{figure}

The vacuum optomechanical coupling rate, $g_{\text{o}}$, can be determined using several techniques. The most widespread and the one used here employs a known frequency modulation as a calibration signal to which the thermally-activated mechanical radio-frequency spectrum is compared. This technique is based on the equivalence of the transduction factor between a laser line undergoing a known frequency/phase fluctuation upon a fixed optical resonance and a fixed laser line driving an optical cavity undergoing (optomechanically-induced) frequency fluctuations. Details on the theoretical description of such a scheme can be found in the original article by Gorodetksy et al.~\cite{gorodetksy_determination_2010} or elsewhere~\cite{gavartin_optomechanical_2011,schneider_optomechanics_2019}.

To measure $g_{\text{o}}$ we employ an electro-optic modulator (details in Subsection S4.2) at a voltage $V$ = 0.178 V to modulate the phase of the laser carrier before it couples to any optical cavity mode. The phase of the input signal is then
\begin{equation}
\label{eq:phasemod}
\phi(t)=\omega_{\text{L}}t+\beta\text{cos}(\Omega_{\text{mod}}t)
\end{equation}
where $\beta$ is the phase shift factor, i.e., $\beta=V/V_{\pi}$, and $\Omega_{\text{mod}}$ the modulation angular frequency. The radio-frequency spectrum exhibits this modulation as a peak at $\Omega_{\text{mod}}/2\pi$, in addition to the Lorentzian lines corresponding to the mechanical modes (see Fig.~\ref{fig:OMCouplingExperimental}). Since the calibration is done with the mechanical modes undergoing incoherent Brownian motion, we can treat the different mechanical modes independently and the power spectral density associated to the individual mechanical resonances and to the calibration tone are given by
\begin{subequations}
\begin{align}
S_{\omega\omega}^{\text{m}}(\Omega)&=8g_{o}^2\overline{n}_{\text{th}}\frac{\Omega_{\text{m}}^2}{(\Omega^2-\Omega_{\text{m}}^2)^2+\Gamma_{\text{m}}^2\Omega_{\text{m}}^2}\\
S_{\omega\omega}^{\text{cal}}(\Omega)&=\frac{1}{2}\Omega_{\text{mod}}^2\beta^2\delta(\Omega-\Omega_{\text{mod}})
\end{align}
\end{subequations}
with $\overline{n}_{th}$ the thermal occupancy of the mechanical mode considered. The power spectral density measurable is given by $S_{VV}(\Omega)=\lvert G_{V\omega}(\Omega) \rvert^2 S_{\omega\omega}(\Omega)$, with $G_{V\omega}(\Omega)$ a dispersive transduction factor. The vacuum optomechanical coupling rate can be obtained directly by comparing the areas $\langle V^2\rangle_m = 2g_o^2\overline{n}\lvert G_{V\omega}(\Omega_m) \rvert^2 $ and $\langle V^2\rangle_{cal} = \frac{\Omega_{mod}^2}{2}\beta^2 \lvert G_{V\omega}(\Omega_{mod}) \rvert^2$ beneath both curves as
\begin{equation}
g_{\text{o}}=\frac{\beta\Omega_{\text{mod}}}{2}\sqrt{\frac{1}{\overline{n}_{\text{th}}}\frac{\langle V^2\rangle_{\text{m}}}{\langle V^2\rangle_{\text{cal}}}}\left\lvert\frac{G_{V\omega}(\Omega_{\text{mod}})}{G_{V\omega}(\Omega_{\text{m}})}\right\rvert
\end{equation}

The transduction function $G_{V\omega}(\Omega)$ is a complicated function which depends on the optical drive scheme \cite{gorodetksy_determination_2010}. To avoid any assumption on it other than its slowly evolving nature, the calibration tone is typically set just a few MHz apart from the mechanical mode of interest and one sets $\left\lvert\frac{G_{V\omega}(\Omega_{\text{mod}})}{G_{V\omega}(\Omega_{\text{m}})}\right\rvert\sim1$. In our case, this would require the tuning of the calibration as many times as mechanical modes are present in our system. We have tested the deviation for a number of significant modes and the error we find in the extracted $g_o$ is less than 10\% of its mean value when $\Omega_{mod}$ is changed within a window spanning 150 to 300 MHz. To simplify the measurement, we tune the calibration tone at a fixed frequency and set $\left\lvert\frac{G_{V\omega}(\Omega_{\text{mod}})}{G_{V\omega}(\Omega_{\text{m}})}\right\rvert=1$ for all modes. We fit the low-power radio-frequency spectra of the Anderson-localized optical modes with a sum of Lorentzian line-shapes baseline fixed by the measured spectrum in spans of around 10 MHz to limit the number of observed modes. As an initial guess for the mechanical frequency, we use the calculated frequencies shown in Fig.~\ref{fig:slot78OM}b. The calibration tone is fixed at $\Omega_{\text{mod}}/2\pi$ = 230 MHz and fitted with a Gaussian line-shape, as it is limited by the chosen electronic spectrum analyzer (ESA) resolution bandwidth (RBW) which is set to 10 kHz. An example of this analysis is shown in Fig.~\ref{fig:OMCouplingExperimental}, where the light-green curves correspond to individual Lorentzian fits and the blue curve to their sum. The vacuum optomechanical coupling extracted for two of the observed peaks are $g_{\text{o}}/2\pi = 68.19 \pm 0.9$ kHz and $g_{\text{o}}/2\pi = 41.3 \pm 1.4$ kHz where the errors are given by the fit error for both the mechanical and calibration signals. Following this approach, we can measure the vacuum optomechanical coupling rate for a large number of mechanical and Anderson-localized modes and extract the probability distribution function of $g_{\text{o}}/2\pi$ for varying air-slot width $s$ (shown in Fig.3c in the main text).

\end{widetext}

\end{document}